\documentclass[12pt]{iopart}

\usepackage{lscape,graphicx}
\usepackage{amssymb}
\usepackage{xpatch}
\usepackage{hyperref}
\usepackage{color}
\usepackage{iopams}

\newcommand{\RED}[1]{{\color{red} #1}}
\newcommand{\beq}{\begin{equation}}
\newcommand{\eeq}{\end{equation}}
 
\newcommand{\drm}{\mathrm{d}}

\begin{document}

\title{A lower limit for Newtonian-noise models of the Einstein Telescope}

\author{Jan Harms$^{1,2}$, Luca Naticchioni$^{3}$, Enrico Calloni$^{4,5}$, Rosario De Rosa$^{4,5}$, Fulvio Ricci$^{3,6}$, Domenico D'Urso$^{7,8}$}
\vskip 1mm
\address{$^{1}$Gran Sasso Science Institute (GSSI), I-67100 L’Aquila, Italy}
\address{$^{2}$INFN, Laboratori Nazionali del Gran Sasso, I-67100 Assergi, Italy}
\address{$^{3}$INFN, Sezione di Roma,  I-00185 Roma, Italy}
\address{$^{4}$Universit\`a degli studi {\it Federico II} Napoli, I-80126 Napoli, Italy}
\address{$^{5}$INFN, Sezione di Napoli, I-80126 Napoli, Italy }
\address{$^6$Universit\`a degli studi di Roma {\it Sapienza}, I-00185 Roma, Italy}
\address{$^{7}$Universit\`a  degli studi di Sassari, I-07100 Sassari, Italy}
\address{$^{8}$INFN, Laboratori Nazionali del Sud, I-95123 Catania, Italy }
\vskip 1mm

\ead{jan.harms@gssi.it}

\vspace{10pt}

\begin{abstract} 
The Einstein Telescope (ET) is a proposed third-generation gravitational-wave (GW) underground observatory. It will have greatly increased sensitivity compared to current GW detectors, and it is designed to extend the observation band down to a few Hz. At these frequencies, a major limitation of the ET sensitivity is predicted to be due to gravitational fluctuations produced by the environment, most importantly by the seismic field, which give rise to the so-called Newtonian noise (NN). Accurate models of ET NN are crucial to assess the compatibility of an ET candidate site with the ET sensitivity target also considering a possible reduction of NN by noise cancellation. With NN models becoming increasingly complex as they include details of geology and topography, it is crucial to have tools to make robust assessments of their accuracy. For this purpose, we derive a lower bound on seismic NN spectra, which is weakly dependent on geology and properties of the seismic field. As a first application, we use the lower limit to compare it with NN estimates recently calculated for the Sardinia and Euregio Meuse-Rhine (EMR) candidate sites. We find the utility of the method, which shows an inconsistency with the predictions for the EMR site, which indicates that ET NN models require further improvement.
\end{abstract}

%
%
%
%
%

\section{Introduction}

The second generation of ground-based interferometric gravitational-wave detectors Advanced LIGO \cite{LSC2015} and Advanced Virgo \cite{AcEA2015} started the era of gravitational-wave (GW) astronomy with the first observations of coalescing massive objects like binary black holes (BBH), binary neutron stars (BNS) and neutron-star black-hole binaries (NSBH), listed in the GW transient catalogues GWTC-1 \cite{AbEA2019}, GWTC-2 \cite{AbEA2021} and GWTC-3 \cite{AbEA2021a}. In April 2020, also KAGRA \cite{AkEA2018} joined the GW international network. The sensitivity band of second generation detectors spans from 10\,Hz to 10\,kHz. The goal of the proposed next-generation detectors Einstein Telescope (ET) \cite{PuEA2010,ET2011,ET2020} and Cosmic Explorer (CE) \cite{ReEA2019,EvEA2021} is to achieve an improvement of sensitivity in this band by about a factor ten.
Furthermore, the observation band of ET will extend down to 2\,Hz. There are strong scientific reasons for pushing the low frequency limit of an interferometric antenna below 10\,Hz \cite{MaEA2020}. Intermediate-mass BBHs can be observed to much higher redshift and with larger total masses. Warnings and sky location of upcoming BNS mergers can be provided earlier in the inspiral phase to better prepare electromagnetic observatories for multi-messenger studies of these sources \cite{ChEA2018}. Many known radiopulsars are observed below 10 Hz. Moreover, young pulsars ($\lesssim 10^7$ yr) spin at low frequencies. These are the neutron stars which most likely deviate significantly from axisymmetry and their emission of continuous GW will give information of their equation of state.

The Einstein Telescope is the European proposal for this new generation of gravitational wave detectors. ET will be a 10\,km-scale multi-interferometer observatory, which will study the Universe in the entire GW spectrum accessible from Earth. This target will be possible thanks to the xylophone configuration of each detector, composed of a pair of complementary interferometers, one optimized for observations at low frequencies and the other with a peak sensitivity at higher frequencies. To reach their target peak sensitivity, the test masses of low-frequency interferometers must be cooled down to 10-20\,K. Moreover, to expand the observation band at frequencies below 10\,Hz, the detectors must be installed in an underground infrastructure, reducing in this way the impact of seismic noise and Newtonian Noise (NN) produced by seismic waves and the atmosphere \cite{BeEA2011,Har2019,BaHa2019}. 

Models of NN are increasingly complex as they incorporate an increasing amount of site-specific information like topography, geology, and observed or modeled inhomogeneities of seismic fields \cite{AnHa2020,BaEA2021}. Systematic and significant numerical errors can affect the results, and one needs to develop methods to constrain simulation results. In this article, we take a first step by providing a lower NN limit, which is a robust benchmark to test results of more complicated simulations since it depends negligibly on geology and properties of the seismic field \footnote{The data used in this paper, and the Python code used to produce the plots are available at the ET site repository \url{https://etrepo.df.unipi.it:8000/}. Accounts can be requested by contacting the authors.}. 

In section \ref{sec:NNmodel}, we recall important aspects of seismic NN modeling for ET. We report in section \ref{sec:seismic} the most recent seismic noise measures obtained with broadband high-class triaxial seismometers installed at about 250\,m depth in boreholes 
at the two ET candidate locations, the Euregio Meuse-Rhine (EMR) site and the Sardinia (Sos Enattos) site. Finally, in section \ref{sec:NNlimit}, we calculate lower bounds on seismic NN for the two sites based on seismic-noise measurements and confront these lower bounds with previously published NN estimates for the two sites.

\section{Seismic NN modeling}
\label{sec:NNmodel}

Modeling of NN has made significant progress over the past years driven by the development of a NN cancellation system as part of an upgrade of the Advanced Virgo detector and for the estimation of NN as part of the evaluation of ET candidate sites \cite{AnHa2020,BaEA2020,AlEA2021,BaEA2021}. Models have been improving by incorporating an increasing amount of observed properties of seismic fields and by including more details in numerical simulations about the local geology and topography. Given the importance of NN estimates for the evaluation of an ET candidate site \cite{BaHa2019,AmEA2020}, a lot of effort is spent to achieve good accuracy with NN models and to understand their limitations. 

Newtonian-noise models inherit the complexity of seismic fields, local geology and topography. When it comes to underground GW detectors like Einstein Telescope, another challenge is to obtain reliable NN estimates based on data from surface seismometers and only a few underground seismometers (borehole installations of seismic broadband sensors are very expensive). The question arises how the accuracy of ET NN models can be assessed under these circumstances. The purpose of this section is to provide insight into how difficult it is to model the different contributions to seismic NN. We also provide a lower limit to the NN spectrum, which only depends on the seismic spectra measured at the test masses of the underground detector. 

\subsection{Simple analytical approach}

A fully general equation to calculate the gravitational fluctuations produced by the displacement field can be calculated from the continuity equation, which states that density perturbations inside the elastic medium are connected to the seismic displacement field $\vec\xi(\vec r,t)$ according to
\beq
\delta \rho(\vec r,t) = - \nabla \cdot \left(\rho(\vec r\,)\vec\xi(\vec r,t)\right),
\eeq
We understand this equation to be linearized with respect to the displacement field, which means that we can use a static mass density $\rho(\vec r\,)$ of the medium inside the divergence. Inserting this term into the expression for the Newtonian gravity potential, we obtain
\beq
\delta\phi(\vec r_0,t) = G\int \drm V  \frac{\nabla\cdot\left(\rho(\vec{r}\,)\vec\xi(\vec r,t)\right)}{|\vec r - \vec r_0|},
\eeq
where $\vec r_0$ can be taken as the location of a (free) test mass. Integration by parts (omitting the surface terms since we can describe any medium, even of finite size as an inhomogeneous medium of infinite size) and applying a gradient operation with respect to the position vector $\vec r_0$, we directly obtain
\beq
\delta\vec a(\vec r_0,t) =  G\int\drm V\rho(\vec r\,)\frac{1}{|\vec r-\vec r_0|^3}\left(\vec\xi(\vec r,t)-3(\vec e_{\delta r}\cdot\vec\xi(\vec r,t))\vec e_{\delta r}\right),
\label{eq:dipoleacc}
\eeq
where $\vec e_{\delta r}\equiv (\vec r-\vec r_0)/|\vec r-\vec r_0|$. This equation can be understood as the sum over a \emph{bulk} term (from the compression of ground material) and of \emph{surface} and interface terms (from seismic displacement along density gradients) \cite{Har2019}. Knowledge of $\rho(\vec r\,)$ and $\vec\xi(\vec r,t)$ will never be complete, and one must proceed with approximations.

Simple theoretical NN models can all be cast into the form
\beq
S(\delta a;f)=(\kappa G\rho)^2 S(\xi;f).
\label{eq:specNN}
\eeq
Here, $\rho$ represents the density of the ground averaged over a sufficiently large surface area or ground volume, and $S(\delta a;f),\, S(\xi;f)$ are the power spectral densities of gravitational acceleration and ground displacement, respectively. The parameter $\kappa$ is characteristic for the type of seismic wave producing NN. Its approximate value is\cite{Har2019}:
\begin{itemize}
    \item $\kappa\approx 8$, for compressional waves and underground test mass;
    \item $\kappa\approx 4$, for shear waves and underground test mass;
    \item $\kappa\approx 4$ -- 6, for Rayleigh waves and surface test mass.
\end{itemize}
A few assumptions have to be made so that these simple estimates of the total NN are approximately valid:
\begin{enumerate}
    \item The distance between test mass and the floor beneath it is significantly smaller than a (reduced) seismic wavelength $\lambda/(2\pi)$;
    \item The geology can be approximated as homogeneous and in the case of a test mass above surface, the topography can be approximated as flat;
    \item In the case of an underground test mass, its depth needs to be at least a seismic wavelength.
\end{enumerate}
If condition (i) is not fulfilled, then NN can be strongly suppressed. For example,  the contribution of surface displacement to NN in underground detectors is negligible \cite{BeEA2011,BaHa2019,AmEA2020}, and in addition a recess constructed under a test mass can lead to substantial reduction of NN from the seismic surface waves \cite{HaHi2014,SHH2020,HaEA2021,SiEA2021}. If condition (ii) is not fulfilled, then one is mostly forced to abandon analytical modeling and directly employ numerical simulations. However, setting up a numerical simulation that correctly recovers all the relevant properties of such an inhomogeneous field is very challenging \cite{AnHa2020,BaEA2021}. Finally, condition (iii) for underground test masses ensures that NN from shear and compressional waves is not significantly affected by the presence of the surface. 

\subsection{Projections in NN models}
Another important aspect of NN modeling is best described in the context of efficient NN cancellation. It concerns the correlations between gravitational acceleration and seismic variables. In equation (\ref{eq:specNN}), it is not clarified where in respect to the test-mass location the seismic spectral density is measured, and along which displacement direction. Of course, from equation (\ref{eq:dipoleacc}), one should not even expect that a seismic displacement at one location fully determines NN spectra. Generally, if we wanted to set up a general relationship between gravitational acceleration and seismic displacements $\xi_k^{x_k}$ at $k=1,\ldots,N$ locations along arbitrary displacement directions $x_k$, we could write it in frequency domain as
\beq
\delta \tilde a_0(f) =  \sum\limits_{k=1}^NF_k(f)\tilde \xi_k^{x_k}(f).
\label{eq:discrNN}
\eeq
All NN models are either described in this form, or by an integral like equation (\ref{eq:dipoleacc}). If presented in the context of observed gravitational acceleration and seismic displacements, the coefficients $F_k(f)$ are also known as Wiener filter. In some sense, a Wiener filter can be understood as a linear NN model, but we will explain a subtle but crucial difference between Wiener filters and most published analytical models. In fact, all published analytical models with finite $N$ are for $N=1$. Why can analytical models be so simple, while Wiener filters even for homogeneous fields are typically constructed for a large number $N$ of measurements  \cite{BeEA2011,DHA2012,CoEA2016a,BaHa2019,HaEA2020}? It has to do with \emph{wave polarization} and with \emph{projections}:
\begin{itemize}
    \item Since the numerical factor $\kappa$ in equation (\ref{eq:specNN}) depends on the polarization of the seismic wave, it is clear that the model coefficients $F_k(f)$ in equation (\ref{eq:discrNN}) must depend on polarization as well. Polarization information is not generally available in measurements of ambient seismic fields, which poses a strong limitation to Wiener filtering whenever more than one wave polarization produces significant NN \cite{BaHa2019}. In analytical models, equations can of course be set up for each polarization separately.
    \item Analytical relations between seismic displacement and gravitational acceleration often (not always) contain a projection-dependent factor, i.e., a factor that depends on the propagation direction of a seismic wave with respect to the modeled direction of gravitational acceleration \cite{Har2019}. Since in an ambient seismic field propagation directions of waves are not always the same, the projection factor cannot appear in Wiener filters, which instead must remain an accurate NN model based on average properties of the seismic field if it is meant to achieve efficient NN cancellation. 
\end{itemize}
Now, it turns out that understanding projections is of crucial interest for NN cancellation \cite{PaHa2016,HaVe2016}, and it guides us to provide extremely simple, accurate analytical models for certain types of seismic fields. Let us illustrate this point for a simple example of gravitational acceleration produced by shear and compressional waves \cite{HaEA2009b}:
\beq
\delta\vec a(\vec r_0,t) =  \frac{4\pi G\rho}{3}\left(2\vec\xi^{\;\rm P}(\vec r_0,t)-\vec\xi^{\;\rm S}(\vec r_0,t)\right).
\label{eq:bodyacc}
\eeq
The equation holds for all seismic fields in infinite, homogeneous media of density $\rho$ that can be decomposed into compressional and shear content $\xi^{\rm P},\,\xi^{\rm S}$, e.g., homogeneity and isotropy of the seismic field are not required. However, the test mass affected by the gravitational fluctuation must be located in a sufficiently small cavern (see section \ref{ssec:lowerlimit}). It is straight-forward to derive this equation for a test mass located in a spherical cavern starting from equation (\ref{eq:dipoleacc}) \cite{HaEA2009b}. The model contains the bulk contribution as well as the contribution from displaced cavern walls. It was heralded as a potential breakthrough for the design of NN cancellation systems for underground detectors:
\begin{itemize}
    \item It states a \emph{local} relation between seismic displacements and gravitational acceleration at $\vec r_0$;
    \item This exact, analytical relation between gravitational acceleration and seismic displacement does not depend on propagation directions of seismic waves.
\end{itemize}
As a consequence, it would be possible to set up a Wiener filter from a local measurement for efficient NN cancellation. In the meantime, the optimism has faded since there is still no known local measurement that would provide the compressional and shear displacements leaving us with a scenario for ET of hundreds of seismometers deployed in boreholes to achieve modest cancellation of NN from body waves \cite{BaHa2019}. However, the equation remains a powerful tool for NN modeling.

\subsection{Model uncertainties and a lower limit of NN}
\label{ssec:lowerlimit}
It is clear that simplified treatment and partial knowledge of geology and topography result in NN modeling errors. Typically, these can be neglected since the NN models are not very sensitive to their parameters, e.g., displacement and density enter linearly into the model. A counter-example is the modeling of underground NN from Rayleigh waves. Here, the wave speed enters exponentially into the NN model, and uncertainties in wave dispersion and/or geology can make a big impact on Rayleigh NN predictions \cite{BeEA2012,AmEA2020,ET2020,SiEA2021}. It is exactly this exponential dependence that is crucial for the strong attenuation of NN towards depth, which is why sophisticated models need to be created and studied to gain confidence in Rayleigh NN predictions for ET \cite{BaEA2021}. According to equation (\ref{eq:bodyacc}), NN from underground displacement is relatively easy to model with all model parameters entering linearly, but one needs to keep in mind that NN predictions from underground displacement need to be quite accurate (see below), which requires us to think more carefully about them as well. We conclude:
\begin{itemize}
    \item Models of NN from Rayleigh waves strongly depend on geology. In the case of horizontally layered geologies, the dependence of Rayleigh NN on depth is almost fully determined by Rayleigh-wave dispersion (with remaining uncertainties coming from linear parameters like ground density). If the geology shows significant lateral heterogeneities, as is usually the case near the surface, then complicated studies are due involving sophisticated numerical simulations matched where possible to surface and underground seismic observations. However, it should be noted that it is easier to cancel Rayleigh NN since it can be done using surface seismic arrays, which somewhat relaxes requirements on model accuracy.
    \item As explained, uncertainties in NN predictions from compressional and shear waves are smaller than for Rayleigh NN, but predictions need to be more accurate since even a modest cancellation of NN from underground seismic fields would be a huge effort \cite{BaHa2019}. A factor 2 error in NN prediction from an underground seismic field can make the difference between evaluating the site as compatible with the ET sensitivity target (naturally or through additional NN cancellation) or as a site where it is difficult to imagine that the ET sensitivity target can be reached during the infrastructure lifetime. The main uncertainties of NN models from underground displacement are (a) the impact of underground infrastructural seismic sources, (b) correlations of NN between test masses, and (c) impact of geology.
\end{itemize}
In the discussion leading to the definition of the NN lower limit for ET, we can neglect Rayleigh-wave NN since it is a contribution without strict lower limit. It depends on the wave dispersion, i.e., on the geology, how much this contribution is attenuated with depth. However, we will be able to show that NN from the underground seismic field is bounded from below (almost) independently of the properties of the seismic field itself and the local geology. 

As mentioned earlier, the NN model of equation (\ref{eq:bodyacc}) contains a bulk contribution and another one from cavern-wall displacement. The bulk contribution depends on geology, and for the definition of the lower limit, we will make the conservative assumption that it does not contribute. So, the idea is to only use the cavern-wall contribution to define a lower NN limit,
\beq
\delta\vec a_{\rm low}(\vec r_0,t) = G\rho\int\drm S\, \left(\vec n(\vec{r})\cdot\vec{\xi}(\vec{r},t)\right)\frac{\vec r-\vec r_0}{|\vec r-\vec r_0|^3}=-\frac{4\pi G\rho}{3}\vec\xi(\vec r_0,t),
\label{eq:lowerNN}
\eeq
which holds precisely like this (with numerical factor $4\pi/3$) only for cubes and spheres as cavern shapes, and possibly other point-symmetric caverns, and the test mass must be placed at its symmetry point. We discuss the impact of cavern shape and test-mass placement on the cavern-wall NN in section \ref{sec:smallcav}.

Compared to equation (\ref{eq:bodyacc}), the lower limit does not require a separation of polarizations. Each wave type --- not only compressional waves and shear waves, but also residual underground displacement from Rayleigh waves --- produces cavern-wall NN according to the same model. One can therefore simply consider the total underground displacement, which is the quantity observed by an underground seismometer, to calculate the lower limit. In this equation, $\rho$ is the density of rock at the cavern, and $\vec n(\vec{r})$ is the normal vector to the cavern wall. Let us summarize the assumptions going into this equation:
\begin{itemize}
    \item The density $\rho$ at the cavern does not change significantly over the extent of the cavern, which can be expected since ET caverns are to be constructed in good quality rock \cite{AmEA2020,ET2020}. 
    \item The cavern hosting the test mass must be sufficiently small. Here, the cavern size is to be compared to the length scale, over which seismic displacement changes. If all seismic waves are from distant sources, and if there is no significant seismic scattering in the medium around the cavern, then distance $r$ from test mass to cavern wall must obey $r\ll\lambda_{\rm S}/(2\pi)$, where $\lambda_{\rm S}$ is the shear-wave length. However, if there are local seismic sources, especially those located inside the cavern as part of the detector infrastructure, or if there is strong seismic scattering in the medium around the cavern, then the cavern might have to be smaller for the lower limit to hold. Assuming a typical shear-wave speed of 3\,km/s at a few 100\,m depth, the reduced wavelength is 50\,m at 10\,Hz, which is significantly larger than the $\sim 10\,$m distance between ET-LF input test masses (ITMs) and cavern walls according to current infrastructure plans \cite{ET2020}. However, with a planned cavern height of 30\,m extending over a length of 175\,m, it is recommended to confirm with numerical simulations that the impact of cavern dimension on NN can indeed be neglected. Note that the cavern dimensions of end test masses (ETMs) are much smaller than for the ITMs, and the ETM NN lower limit is robust in this respect.
    \item When adding contributions of different test masses to obtain the total ET NN lower limit, we assume that gravity fluctuations are uncorrelated between test masses. This is certainly the case between test masses forming an interferometer arm, which will be separated by 10\,km in ET. Instead, the two ITMs of an ET-LF interferometer will be separated by about 100\,m according to current plans \cite{ET2020}. There might well be significant gravitational correlation between these two test masses. According to equation (\ref{eq:bodyacc}), gravitational correlations between two test masses are determined by the seismic correlations between the same two points. Measurements at the Sanford Underground Research Facility showed that seismic correlations are small above 4\,Hz for a seismometer separation of 190\,m, even if displacements are monitored along the same directions \cite{CoEA2018b}. New underground measurements are required to obtain better estimates of NN correlations between the ET-LF ITMs in the 3\,Hz to 10\,Hz band.
\end{itemize}

\subsection{Additional notes on the small-cavern limit}
\label{sec:smallcav}
Let us shed more light on the small-cavern approximation used for equation (\ref{eq:lowerNN}). It might seem surprising that the gravitational acceleration points along the displacement direction of the seismic field. After all, the integrand in equation (\ref{eq:dipoleacc}) contains a contribution that is not parallel to $\vec\xi$. It is clear though that the result of the integration must be expandable in terms of physical vectors independent of the chosen coordinate system. The vectors available here are $\vec\xi(\vec r_0)$ and the normal vectors to the cavern wall $\vec n(\vec r)$. Equation (\ref{eq:lowerNN}) was initially calculated for a spherical cavern, where terms proportional to the cavern normal vector vanish (actually, this is true for all spherical cavern sizes). When considering arbitrary shapes of caverns, could it be that the gravitational acceleration is not parallel to ground displacement in equation (\ref{eq:lowerNN})?

One needs to be careful when answering to this question. If one placed a seismometer on the ground of a cavern to measure vertical ground displacement, then it would generally show correlations with the gravitational fluctuations along the horizontal direction. So, one \emph{could} construct an analytical model that includes vectors normal to the cavern walls. However, we are looking for the simplest (still accurate) NN model. Crucial for why gravity acceleration is parallel to ground displacement for all cavern shapes is that in the small-cavern limit the integrand in equation (\ref{eq:lowerNN}) becomes independent of the distance between test mass and cavern wall. This is easy to derive by making use of the fact that in the small-cavern limit the displacement vector has the same value $\vec\xi(\vec r_0,t)$ over the entire cavern wall. In this case, the closed surface integral leaves the seismic displacement $\vec\xi(\vec r_0)$ as sole physical direction for the gravitational acceleration. Mathematically, this result is easiest to derive by expanding the integrand of the small-cavern approximation of the surface integral, 
\beq
\delta\vec a_{\rm low}(\vec r_0,t) = G\rho\int\drm\Omega\, \left(\vec n(\theta,\phi)\cdot\vec{\xi}(\vec{r}_0,t)\right)\vec e_{\delta r},
\label{eq:smallcav}
\eeq
into vector surface spherical harmonics and using equation (253) in \cite{Har2019} with $\vec\xi(\vec{r}_0,t)$ as polar direction of a spherical coordinate system. The integral is over the angular coordinates $\drm\Omega = \drm\phi\,\drm\theta\sin(\theta)$ of a spherical coordinate system with origin at $\vec r_0$. 

The remaining question is about the coupling factor $4\pi/3$ in equation (\ref{eq:smallcav}). We already mentioned that it depends on cavern shape and test-mass location inside the cavern. We summarize the values of this factor (obtained analytically or by numerical integration) for several interesting cases in table \ref{tab:caverns}.
\begin{table}[ht!]
    \centering
    \begin{tabular}{p{5cm}|c|c}
         \bf Cavern shape & \bf Test-mass location & \bf Coupling factor \\
         \hline
         Spherical (as used in \cite{AlEA2021,BaEA2021}) & arbitrary & $4\pi/3$\\
         \hline
         Cube & 3D center & $4\pi/3$\\
         & 2D center on floor & 2.8 \\
         \hline 
         Half sphere & 2D center on floor & 2.1 \\
         \hline
         \raggedright Cuboid elongated along horizontal (side-length ratio: 2:1:1) & 2D center on floor & 1.3
    \end{tabular}
    \caption{Summary of cavern-wall NN coupling factors for different cavern shapes and test-mass locations valid in the small-cavern approximation. The dependence of total NN on cavern-wall coupling factors is less since a major NN contribution underground is expected to come from rock compression.}
    \label{tab:caverns}
\end{table}
The coupling factors for cavern-wall NN in the small-cavern approximation can be significantly smaller than $4\pi/3$ depending on cavern shape and test-mass location. This however is only relevant for the lower seismic NN bound that relies entirely on the cavern-wall contribution. \emph{The total seismic NN with a major bulk contribution is less affected by cavern shape and test-mass location provided that the small-cavern approximation holds.} Based on the coupling factors shown in table \ref{tab:caverns}, it is evident that using spherical caverns in NN models leads to the most stringent confrontation between lower limit and results of complex numerical simulations, since the lower limit is generally smaller for other cavern shapes. In fact, we propose that all NN models, even those based on more realistic cavern shapes, should also be run with small spherical caverns to set the highest possible lower bound on the more complicated NN contributions that depend on geology, topography, and details of the seismic field.
\begin{landscape}
\begin{table}[ht]
\centering
\includegraphics[width=1.4\textwidth]{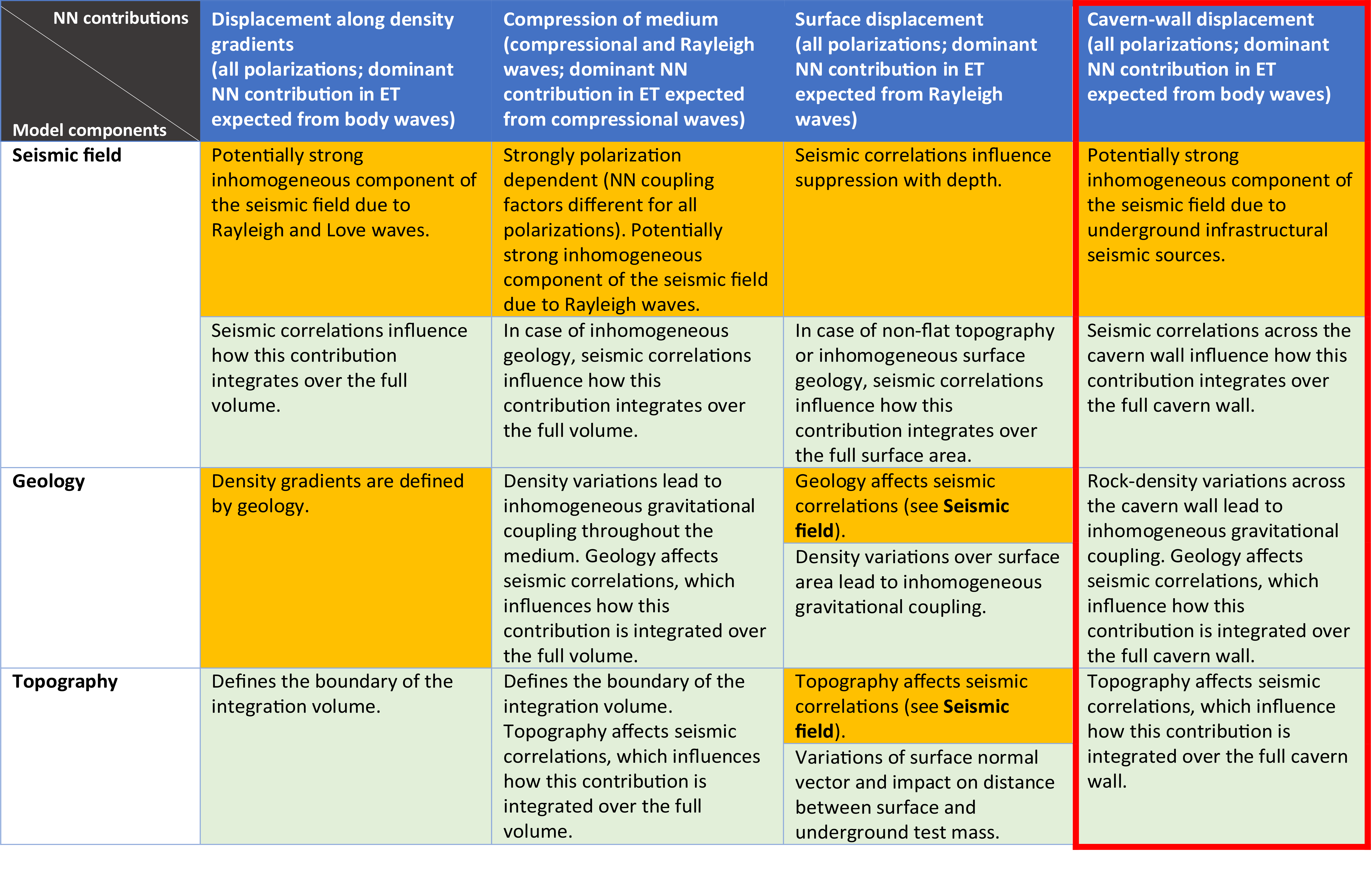}
\parbox{1.3\textheight}{\caption{Summary table of seismic NN contributions and model components. The NN contributions can be assumed to be approximately uncorrelated being dominated by different types of seismic waves, or due to well separated locations where these contributions are produced, or due to differences in gravitational coupling mechanisms. The green fields in this table indicate a weak or negligible influence on the respective NN contribution, the orange fields indicate a strong or potentially strong influence. The lower NN limit presented in this paper is solely based on the cavern-wall contribution in the last column.}\label{tab:summary}}
\end{table}
\end{landscape}

In conclusion, with some points that still need to be analyzed like the impact of ET-LF ITM cavern size on NN and potential NN suppression by partial correlation of NN between the two ITMs of an interferometer, the lower seismic NN limit presented here serves as a useful reference for consistency checks of total NN estimates. As defined in equation (\ref{eq:lowerNN}), it applies to all published analyses since neither the ITM NN correlation nor the ET ITM cavern dimension have been taken into consideration in past analyses, and the cavern shape was always assumed to be spherical. 

As a final note, we want to point out that while the definition of the lower limit itself only depends on the stated assumptions so far, there is a general underlying assumption that the different seismic NN contributions summarized in table \ref{tab:summary} are at best weakly correlated. In the past, people used the shell theorem (the case of perfectly anti-correlated NN contributions from the two surfaces of the shell) to conclude that simple cavern-wall displacement cannot produce NN. The argument is that if you interpret an infinite medium with spherical cavern as a shell with infinite outer radius, uniform motion of the medium (which still leads to cavern-wall displacement), should not produce any change in gravity. However, this example has no direct relevance to the problem of cavern-wall NN. First, the calculation of gravitational forces inside caverns of an infinite medium is an ill-defined problem. The infinite medium can be constructed as a limit not only of spherical shells, but also of spherical caverns inside a cube, etc, with different conclusions about the force. More importantly though, while the displacement of a cavern wall can be considered uniform in the case of a small cavern, the outer boundary (whatever shape it is) will show non-uniform displacement consistent with the solution of the gravitoelastic equations. This has important consequences especially in the case of random seismic fields, which is the relevant case to understand for ET NN modeling. The following argument is easiest to explain in the small-cavern approximation, but it is applicable to the general case. The wall of a small cavern is displaced uniformly, i.e., it depends on the seismic displacement at a single point. All other contributions are the consequence of integrals over extended fields (as volume or surface integrals) with contributions from uncorrelated parts of the seismic field within the integration region. It is clear that this can at best lead to a weak correlation between cavern-wall NN with any other NN contribution. Now, let us imagine anyway that there is a mechanism that produces negative correlation between cavern-wall NN and another NN contribution, which could potentially lead to a total NN spectrum below the lower limit. Even for a mild NN suppression by a factor 1.5, one needs a coherence \cite{Har2019}
\beq
\gamma(f)=\frac{\mathcal R(\langle x(f)y^*(f)\rangle)}{\sqrt{\langle |x(f)|^2\rangle\langle |y(f)|^2\rangle}}\in [-1,1]\leq -0.75,
\eeq
where $\langle x(f)y^*(f)\rangle$ is the cross power spectral density of two NN contributions, and $\langle |x(f)|^2\rangle,\,\langle |y(f)|^2\rangle$ the power spectral densities. This level of negative correlation is inconceivable for seismic NN from random seismic fields, which should be intuitively clear when taking into account that the cross-spectrum relevant here is between a NN contribution that depends on the value of the seismic field at only one point and other NN contributions integrated over extended regions of the seismic field with contributions from uncorrelated parts and displacement directions of the seismic field (for analyses of seismic correlations, see for example \cite{CoEA2018a,CoEA2018b,TrEA2019}).

\section{Seismic noise measurements}
\label{sec:seismic}
As explained in the previous section, the knowledge of the seismic field plays a crucial role in the estimation of the expected NN. For this purpose, several seismic stations were installed at the two candidate sites, including temporary sensor arrays and permanent seismometer installations \cite{Naticchioni_2020,etsar2, etterz}. The best option to achieve a reliable underground seismic measurement is through broadband tri-axial seismometers in borehole installations \cite{AmEA2020}. Considering the proposed design of the ET infrastructure, a representative depth of 250\,m from the surface level was adopted as target for these measurements. At the EMR site, the border region between Netherlands, Belgium and Germany, a first borehole was excavated close to the village of Terziet in the southern countryside of Limburg province, and it is equipped with a Streckeisen STS-5A seismometer, placed 250\,m underground, complemented by a Nanometrics Trillium-240 positioned at the surface level \cite{etterz}. Data  from both sensors is sampled at 40 Hz. The Terziet seismic station is operative since late 2019. 

In the Italian island of Sardinia, the other candidate site, two boreholes have been excavated close to two tentative corner locations (named P2 and P3) of a triangle with 10\,km side length, where the first corner is represented by the former Sos Enattos mine \cite{Naticchioni_2020,etsar2}. At both borehole stations, Nanometrics Trillium-SPH2 seismometers are installed about 250\,m underground, complemented by Nanometrics Trillium-120 Horizon seismometers in vault installations at the surface \cite{ET-0426A-21}. Data is acquired by a 24-bit Nanometrics Centaur digital recorder with a sampling frequency of 100 Hz. P2 and P3 stations are operative since September 2021. 

Data from all the permanent borehole stations are collected on a dedicated ET data repository and available for analysis. In the following subsections, we show the probabilistic power spectral densities (PPSD), i.e., histograms of seismic spectra, calculated for Terziet and P2 boreholes, representing the EMR and the Sardinia underground seismic noise level. In the first case, we use one year of data, while for P2 the last three months of 2021, i.e., the entire available dataset from P2 at the time of paper writing. For comparison, we indicate also the new low-noise model (NLNM) and new high-noise model (NHNM) representing the minimum and maximum natural seismic-noise background observed on Earth \cite{Pet1993}. Individual seismic PSDs are calculated using Hann anti-leakage windows and without averaging of PSDs going into the PPSDs. Spectral histograms are computed for the horizontal and vertical components.

\subsection{EMR site}
\begin{figure}[ht!]
\centering\includegraphics[width=0.49\textwidth]{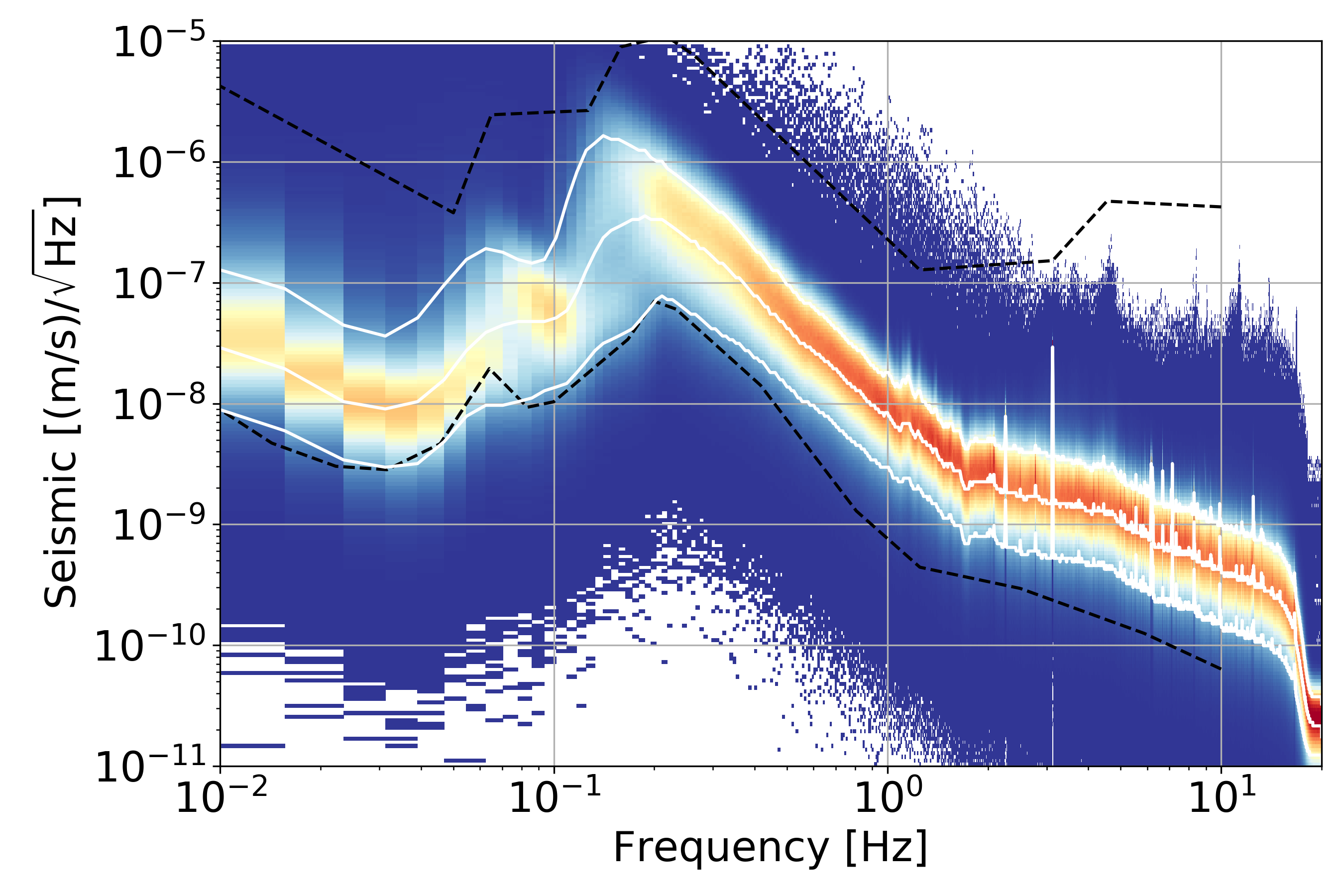}
\includegraphics[width=0.49\textwidth]{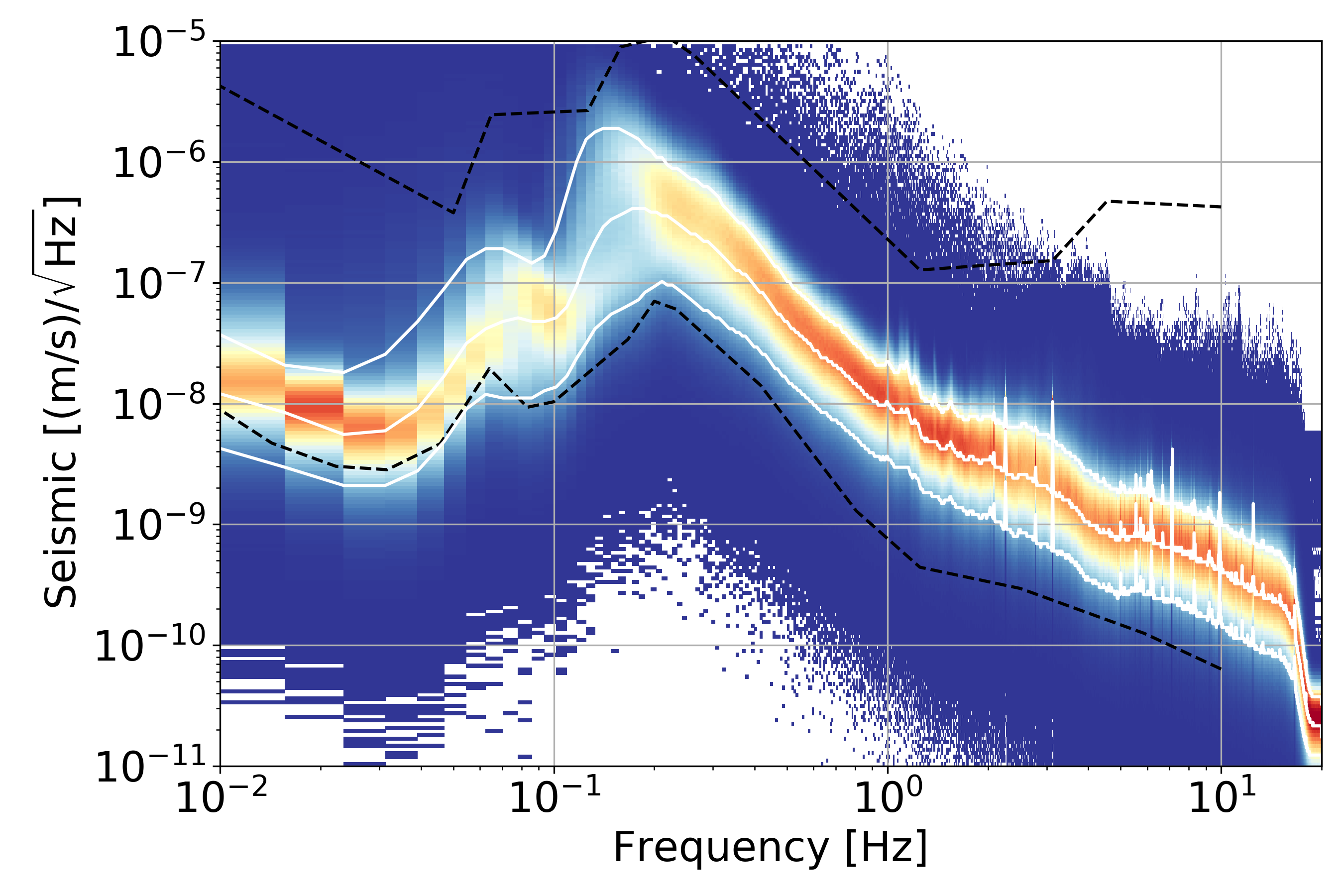}
\caption{Histogram of horizontal (left) and vertical (right) ground motion measured in the Terziet borehole, EMR site, at 250\,m depth (data from September 30, 2019 to September 14, 2020). The dashed curves represent the New Low-noise Model (NLNM) and high-noise model \cite{Pet1993}. White curves are the 10th, 50th and 90th percentiles of the distribution.}
\label{fig:HHEterz}
\end{figure}

The Terziet borehole station is placed at about 152\,m above mean sea level. The borehole seismometer is located at a depth of 250\,m inside the 260\,m-deep borehole \cite{etterz}. We analyzed one year of data from this station, between September 30th, 2019 and September 14th, 2020. In figure \ref{fig:HHEterz}, we show the PPSD of the horizontal and vertical velocities measured at 250\,m of depth in the Terziet borehole.

\subsection{Sardinia site}
\begin{figure}[ht!]
\centering\includegraphics[width=0.49\textwidth]{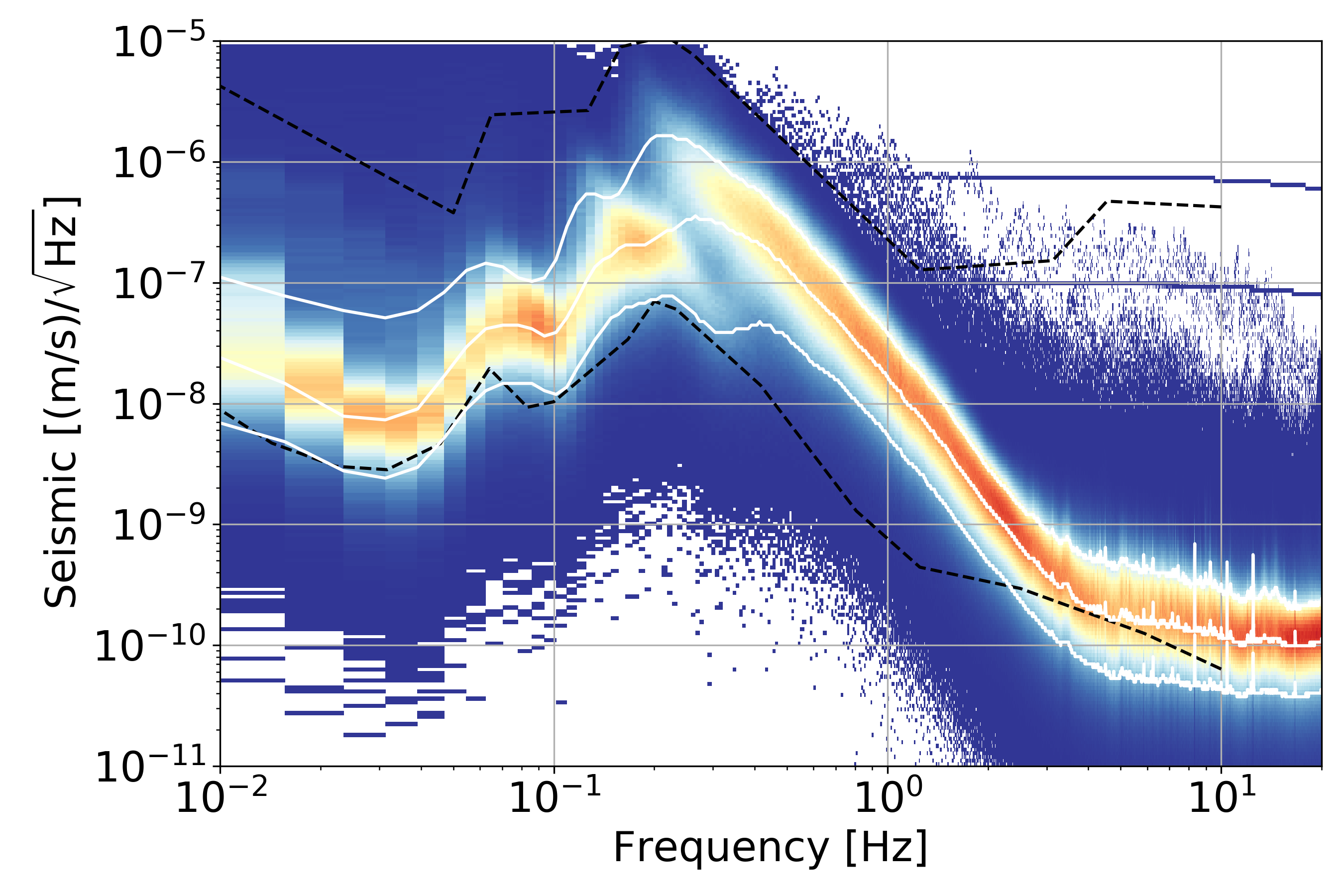}
\includegraphics[width=0.49\textwidth]{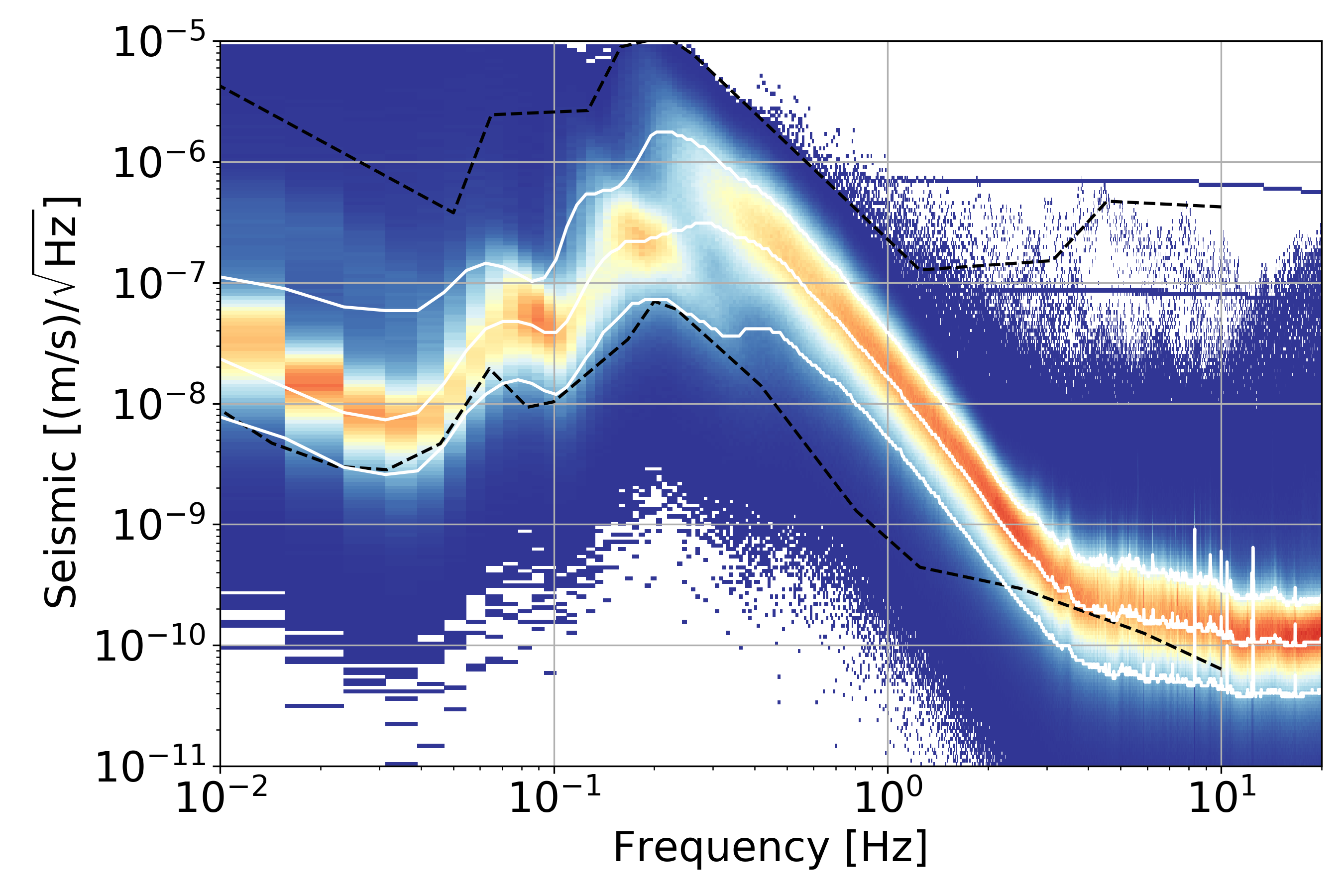}
\caption{Histogram of horizontal (left) and vertical (right) ground motion measured in the P2 borehole, Sardinia site, at 264\,m depth (data from October 1, 2021 to December 31, 2021). The dashed curves represent the New Low-noise Model (NLNM) and high-noise model \cite{Pet1993}. White curves are the 10th, 50th and 90th percentiles of the distribution.}
\label{fig:HHEp2}
\end{figure}

\begin{figure}[ht!]
\centering\includegraphics[width=0.49\textwidth]{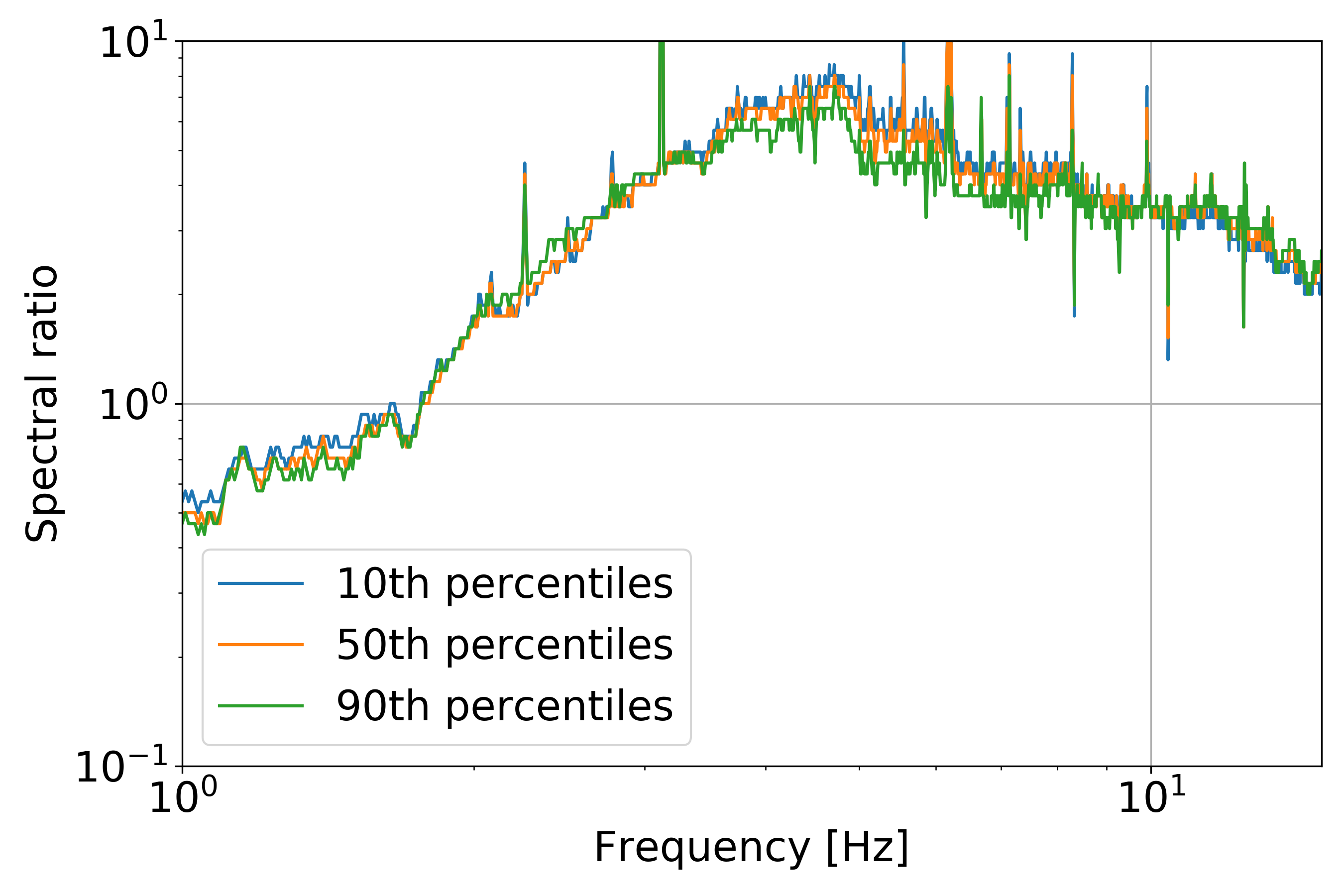}
\includegraphics[width=0.49\textwidth]{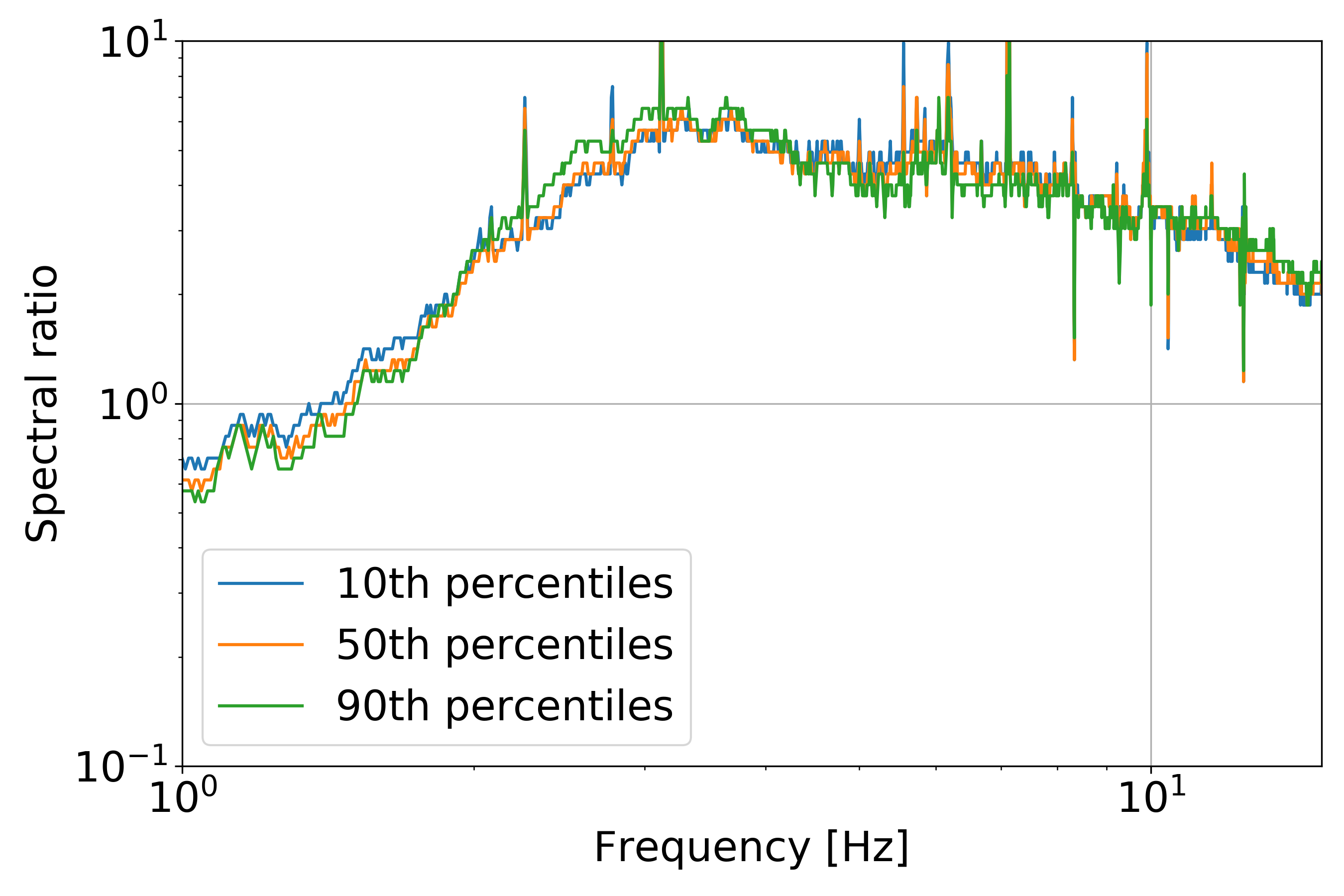}
\caption{Ratio of percentiles of horizontal (left) and vertical (right) ground motion measured in the Terziet borehole at the EMR site relative to the spectrum measured in the P2 borehole at the Sardinian site.}\label{fig:ratio}
\end{figure}

The P2 station is located in the municipality of Bitti, in a private land used for grazing, at about 767\,m above mean sea level. The borehole seismometer, installed inside the 270\,m-deep borehole, is located at a depth of 264\,m, with a residual tilt of the sensor from the vertical direction of $1^{\circ}$ \cite{ET-0426A-21}, within the sensor mass-centering requirements. For this site, we used data limited to the last three months of 2021, since this borehole station is operative since late September 2021. In figure \ref{fig:HHEp2}, we show the PPSD of the horizontal and vertical ground velocities measured at 264\,m of depth in the P2 borehole.
\RED{The} 
ratio of percentiles of seismic spectra measured at the two sites (Terziet borehole relative to P2) are shown for horizontal and vertical ground displacement in figure \ref{fig:ratio}.

\section{NN lower limits}\label{sec:NNlimit}
\begin{figure}[ht!]
\centering\includegraphics[width=0.7\textwidth]{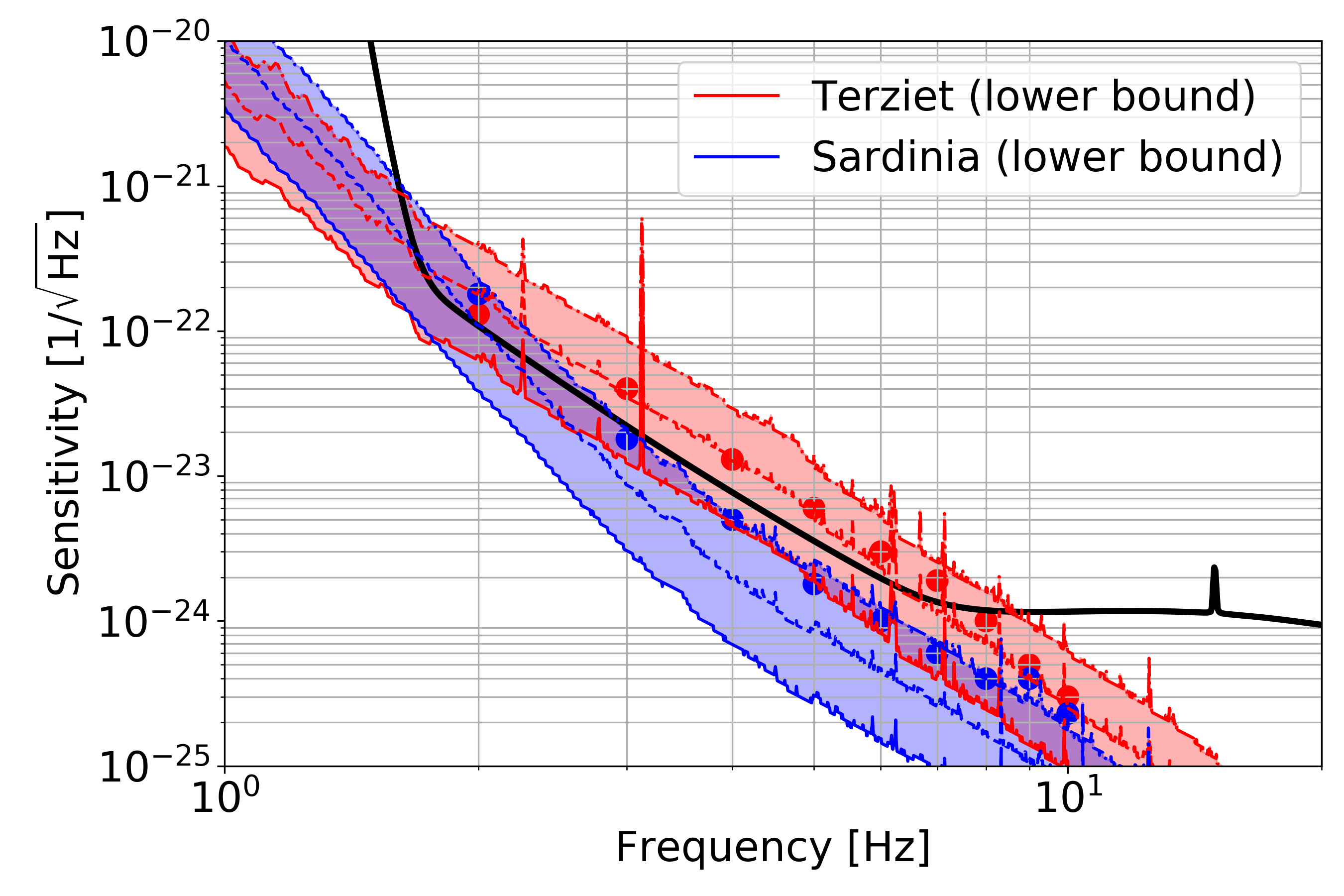}
\caption{The plot shows the lower bounds on seismic NN for the two sites based on the 10th, 50th, and 90th percentiles of the seismic histograms using equation (\ref{eq:lowerNN}) with $\rho=2800\,$kg/m$^3$ for the density of rock at the ET caverns. The ET sensitivity target is shown by the black curve. The Terziet NN estimates at the EMR site published in \cite{BaEA2021}, are shown as red markers for comparison. Limitations in the analysis acknowledged in \cite{BaEA2021} concerning the estimated mode composition of the seismic field might explain why their total NN estimate lies on or below the median of the lower bound presented here, which only includes the cavern-wall contribution. The comparison indicates an inconsistency between the two results. The NN estimates for Sardinia published in \cite{AlEA2021} are shown as blue markers, which lie well above the median of the lower bound. This is to be expected when the model of the total NN predicts a significant contribution from seismic compression of the medium.}\label{fig:NNlimits}
\end{figure}

As motivated in section \ref{sec:NNmodel}, we evaluate the NN produced by the cavern-wall displacement on a test mass, taking into account the total horizontal displacement measured by the borehole seismometers located in EMR and Sardinia.
In figure \ref{fig:NNlimits}, we compare the 10th, 50th and 90th percentiles of the NN at the two sites. Here we assumed a rock density $\rho = 2800\,$kg/m$^3$ for both. Clearly, the lower measured seismic displacement in Sardinia, with respect to the EMR site, reflects also on the NN lower limit especially above 3\,Hz.

The results show that the previously published seismic NN estimates for the Sardinia site lie well above median of the lower NN limit. This does not mean that these estimates are necessarily accurate. In fact, a very simple uniform geology was assumed to calculate the NN estimate in \cite{AlEA2021}, and a conservative assumption was made concerning the mode content of the underground seismic field. However, despite the simplifying assumptions (or in this case rather \emph{because of} the simplifying assumptions), the NN estimates are consistent with the lower bound. 

In comparison, the total NN estimate for the EMR site lies mostly right on, and partially also below the lower bound. Given that the cavern-wall NN is a minor contribution to the total NN, we conclude that we conclude that the EMR NN estimates are too low over part of the NN frequency band. The analysis in \cite{BaEA2021} that led to the NN estimate is sophisticated as it attempts a decomposition of the seismic field into shear waves and compressional-waves from distant sources, and waves without specified polarization from local surface sources. It is when complex models like this are used that the lower NN limit becomes most effective depending only on the seismic spectrum measured underground. We conclude that further work is necessary to improve current NN models.

We want to stress again that especially the NN contribution from underground seismic fields must be modeled with high accuracy. Even in the ideal case of an isotropic seismic field whose two-point correlations do not change in time, it was estimated that a few hundred borehole seismometers would be needed for ET for a factor 2 NN reduction over the entire NN band \cite{BaHa2019}. One might expect that time variation of the seismic field will make the problem even more challenging. In such a situation, even small NN estimation errors can make an important difference. It is then advisable that systematic errors in NN estimates should be taken into account as part of future analyses.

\section{Conclusions and discussion}
The Einstein Telescope, a third-generation GW detector, will expand the observation band down to 2\,Hz. In this region of frequencies, the seismic Newtonian noise (NN) is a main limitation to the targeted sensitivity. A robust modeling of this source of noise for the underground detectors is of paramount importance in the site qualification process and to design efficient cancellation systems. Operating ET in an underground infrastructure will ensure that the NN due to surface displacement will be reduced. Models describing the NN produced by Rayleigh waves depends on the accurate knowledge of the geologic setup. Nevertheless, surface seismic arrays are less costly to deploy and thereby can relax the requirements on the model accuracy. 

Conversely, models accounting for NN due to compressional or shear waves must be more accurate since the cancellation of this contribution is more complex and expensive requiring a large number of borehole seismometer installations. Two terms contribute to the gravitational acceleration of a test mass due to NN: the bulk contribution, and the cavern-wall displacement. The first one might depend significantly on geology. For the lower NN limit presented in this paper, we assumed that the (otherwise dominant) bulk contribution does not contribute. What we win by this rough approximation is that the lower NN limit becomes weakly dependent on local geology, and it is calculated only using the measured horizontal ground displacement at the underground level.

We analyzed the seismic data produced by borehole seismometers located at a depth of $\sim 250\,$m at the two candidate sites, EMR and Sardinia. In the band of interest for ET, i.e., between 2\,Hz and 20\,Hz, the measured horizontal seismic displacement is lower in Sardinia with respect to EMR by up to a factor $\sim 8$ at 5\,Hz. Thus, the computed lower NN limit in Sardinia is smaller by the same ratio than the lower limit of the EMR site. We find the median of the EMR lower limit to lie above the ET sensitivity target by about a factor 1.5 -- 2, while the median of the Sardinia lower NN limit (above 3\,Hz also its 90th percentile) lies below the ET sensitivity target.

The main purpose of the lower NN bound is not to set minimal requirements on a NN cancellation system, which must in the end come out of accurate estimates of the total seismic NN. Instead, we see the lower NN bound as an important reference for future NN models, where increased model complexity might cause larger systematic errors. We found that results of a previous NN study for the EMR site, which relies on a more complex analysis method, lies below the lower NN limit calculated for the EMR site. In order to advance with NN analyses, we need to include an assessment of systematic errors by comparing different approaches and levels of complexity of NN models, with the aim to ultimately define a robust standard for the evaluation of NN.

\ack{The support of {\it Sapienza Grande Progetto di Ateneo 2017, Amaldi Research Center (MIUR program “Dipartimento di Eccellenza”} CUP:B81I18001170001) and INFN is gratefully acknowledged. The authors are also gratefull for the support from the MIUR with the agreements 0021983-06/03/2018 and the PRIN 2017 Research Program Framework no. 2017SYRTCN.
}

\section*{References}
\bibliographystyle{iopart-num}
\bibliography{references}

\providecommand{\newblock}{}
\begin{thebibliography}{10}
\expandafter\ifx\csname url\endcsname\relax
  \def\url#1{{\tt #1}}\fi
\expandafter\ifx\csname urlprefix\endcsname\relax\def\urlprefix{URL }\fi
\providecommand{\eprint}[2][]{\url{#2}}

\bibitem{LSC2015}
Aasi J, Abbott B~P, Abbott R, Abbott T, Abernathy M~R, Ackley K, Adams C, Adams
  T, Addesso P, Adhikari R~X, Adya V, Affeldt C, Aggarwal N, Aguiar O~D, Ain A,
  Ajith P, Alemic A, Allen B, Amariutei D, Anderson S~B, Anderson W~G, Arai K,
  Araya M~C, Arceneaux C, Areeda J~S, Ashton G, Ast S, Aston S~M, Aufmuth P,
  Aulbert C, Aylott B~E, Babak S, Baker P~T, Ballmer S~W, Barayoga J~C, Barbet
  M, Barclay S, Barish B~C, Barker D, Barr B, Barsotti L, Bartlett J, Barton
  M~A, Bartos I, Bassiri R, Batch J~C, Baune C, Behnke B, Bell A~S, Bell C,
  Benacquista M, Bergman J, Bergmann G, Berry C~P~L, Betzwieser J, Bhagwat S,
  Bhandare R, Bilenko I~A, Billingsley G, Birch J, Biscans S, Biwer C,
  Blackburn J~K, Blackburn L, Blair C~D, Blair D, Bock O, Bodiya T~P, Bojtos P,
  Bond C, Bork R, Born M, Bose S, Brady P~R, Braginsky V~B, Brau J~E, Bridges
  D~O, Brinkmann M, Brooks A~F, Brown D~A, Brown D~D, Brown N~M, Buchman S,
  Buikema A, Buonanno A, Cadonati L, Bustillo J~C, Camp J~B, Cannon K~C, Cao J,
  Capano C~D, Caride S, Caudill S, Cavagli{\`{a}} M, Cepeda C, Chakraborty R,
  Chalermsongsak T, Chamberlin S~J, Chao S, Charlton P, Chen Y, Cho H~S, Cho M,
  Chow J~H, Christensen N, Chu Q, Chung S, Ciani G, Clara F, Clark J~A,
  Collette C, Cominsky L, Constancio M, Cook D, Corbitt T~R, Cornish N, Corsi
  A, Costa C~A, Coughlin M~W, Countryman S, Couvares P, Coward D~M, Cowart M~J,
  Coyne D~C, Coyne R, Craig K, Creighton J~D~E, Creighton T~D, Cripe J, Crowder
  S~G, Cumming A, Cunningham L, Cutler C, Dahl K, Canton T~D, Damjanic M,
  Danilishin S~L, Danzmann K, Dartez L, Dave I, Daveloza H, Davies G~S, Daw
  E~J, DeBra D, Pozzo W~D, Denker T, Dent T, Dergachev V, DeRosa R~T, DeSalvo
  R, Dhurandhar S, D{\textasciiacute}{\i}az M, Palma I~D, Dojcinoski G,
  Dominguez E, Donovan F, Dooley K~L, Doravari S, Douglas R, Downes T~P,
  Driggers J~C, Du Z, Dwyer S, Eberle T, Edo T, Edwards M, Edwards M, Effler A,
  Eggenstein H~B, Ehrens P, Eichholz J, Eikenberry S~S, Essick R, Etzel T,
  Evans M, Evans T, Factourovich M, Fairhurst S, Fan X, Fang Q, Farr B, Farr
  W~M, Favata M, Fays M, Fehrmann H, Fejer M~M, Feldbaum D, Ferreira E~C,
  Fisher R~P, Frei Z, Freise A, Frey R, Fricke T~T, Fritschel P, Frolov V~V,
  Fuentes-Tapia S, Fulda P, Fyffe M, Gair J~R, Gaonkar S, Gehrels N,
  Gergely{\textasciiacute} L~{\'{A}}, Giaime J~A, Giardina K~D, Gleason J,
  Goetz E, Goetz R, Gondan L, Gonz{\'{a}}lez G, Gordon N, Gorodetsky M~L,
  Gossan S, Go{\ss}ler S, Gräf C, Graff P~B, Grant A, Gras S, Gray C,
  Greenhalgh R~J~S, Gretarsson A~M, Grote H, Grunewald S, Guido C~J, Guo X,
  Gushwa K, Gustafson E~K, Gustafson R, Hacker J, Hall E~D, Hammond G, Hanke M,
  Hanks J, Hanna C, Hannam M~D, Hanson J, Hardwick T, Harry G~M, Harry I~W,
  Hart M, Hartman M~T, Haster C~J, Haughian K, Hee S, Heintze M, Heinzel G,
  Hendry M, Heng I~S, Heptonstall A~W, Heurs M, Hewitson M, Hild S, Hoak D,
  Hodge K~A, Hollitt S~E, Holt K, Hopkins P, Hosken D~J, Hough J, Houston E,
  Howell E~J, Hu Y~M, Huerta E, Hughey B, Husa S, Huttner S~H, Huynh M,
  Huynh-Dinh T, Idrisy A, Indik N, Ingram D~R, Inta R, Islas G, Isler J~C,
  Isogai T, Iyer B~R, Izumi K, Jacobson M, Jang H, Jawahar S, Ji Y,
  Jim{\'{e}}nez-Forteza F, Johnson W~W, Jones D~I, Jones R, Ju L, Haris K,
  Kalogera V, Kandhasamy S, Kang G, Kanner J~B, Katsavounidis E, Katzman W,
  Kaufer H, Kaufer S, Kaur T, Kawabe K, Kawazoe F, Keiser G~M, Keitel D, Kelley
  D~B, Kells W, Keppel D~G, Key J~S, Khalaidovski A, Khalili F~Y, Khazanov E~A,
  Kim C, Kim K, Kim N~G, Kim N, Kim Y~M, King E~J, King P~J, Kinzel D~L, Kissel
  J~S, Klimenko S, Kline J, Koehlenbeck S, Kokeyama K, Kondrashov V, Korobko M,
  Korth W~Z, Kozak D~B, Kringel V, Krishnan B, Krueger C, Kuehn G, Kumar A,
  Kumar P, Kuo L, Landry M, Lantz B, Larson S, Lasky P~D, Lazzarini A, Lazzaro
  C, Le J, Leaci P, Leavey S, Lebigot E~O, Lee C~H, Lee H~K, Lee H~M, Leong
  J~R, Levin Y, Levine B, Lewis J, Li T~G~F, Libbrecht K, Libson A, Lin A~C,
  Littenberg T~B, Lockerbie N~A, Lockett V, Logue J, Lombardi A~L, Lormand M,
  Lough J, Lubinski M~J, Lück H, Lundgren A~P, Lynch R, Ma Y, Macarthur J,
  MacDonald T, Machenschalk B, MacInnis M, Macleod D~M, Maga{\~{n}}a-Sandoval
  F, Magee R, Mageswaran M, Maglione C, Mailand K, Mandel I, Mandic V, Mangano
  V, Mansell G~L, M{\'{a}}rka S, M{\'{a}}rka Z, Markosyan A, Maros E, Martin
  I~W, Martin R~M, Martynov D, Marx J~N, Mason K, Massinger T~J, Matichard F,
  Matone L, Mavalvala N, Mazumder N, Mazzolo G, McCarthy R, McClelland D~E,
  McCormick S, McGuire S~C, McIntyre G, McIver J, McLin K, McWilliams S,
  Meadors G~D, Meinders M, Melatos A, Mendell G, Mercer R~A, Meshkov S,
  Messenger C, Meyers P~M, Miao H, Middleton H, Mikhailov E~E, Miller A, Miller
  J, Millhouse M, Ming J, Mirshekari S, Mishra C, Mitra S, Mitrofanov V~P,
  Mitselmakher G, Mittleman R, Moe B, Mohanty S~D, Mohapatra S~R~P, Moore B,
  Moraru D, Moreno G, Morriss S~R, Mossavi K, Mow-Lowry C~M, Mueller C~L,
  Mueller G, Mukherjee S, Mullavey A, Munch J, Murphy D, Murray P~G, Mytidis A,
  Nash T, Nayak R~K, Necula V, Nedkova K, Newton G, Nguyen T, Nielsen A~B,
  Nissanke S, Nitz A~H, Nolting D, Normandin M~E~N, Nuttall L~K, Ochsner E,
  O'Dell J, Oelker E, Ogin G~H, Oh J~J, Oh S~H, Ohme F, Oppermann P, Oram R,
  O'Reilly B, Ortega W, O'Shaughnessy R, Osthelder C, Ott C~D, Ottaway D~J,
  Ottens R~S, Overmier H, Owen B~J, Padilla C, Pai A, Pai S, Palashov O,
  Pal-Singh A, Pan H, Pankow C, Pannarale F, Pant B~C, Papa M~A, Paris H,
  Patrick Z, Pedraza M, Pekowsky L, Pele A, Penn S, Perreca A, Phelps M, Pierro
  V, Pinto I~M, Pitkin M, Poeld J, Post A, Poteomkin A, Powell J, Prasad J,
  Predoi V, Premachandra S, Prestegard T, Price L~R, Principe M, Privitera S,
  Prix R, Prokhorov L, Puncken O, Pürrer M, Qin J, Quetschke V, Quintero E,
  Quiroga G, Quitzow-James R, Raab F~J, Rabeling D~S, Radkins H, Raffai P, Raja
  S, Rajalakshmi G, Rakhmanov M, Ramirez K, Raymond V, Reed C~M, Reid S, Reitze
  D~H, Reula O, Riles K, Robertson N~A, Robie R, Rollins J~G, Roma V, Romano
  J~D, Romanov G, Romie J~H, Rowan S, Rüdiger A, Ryan K, Sachdev S, Sadecki T,
  Sadeghian L, Saleem M, Salemi F, Sammut L, Sandberg V, Sanders J~R, Sannibale
  V, Santiago-Prieto I, Sathyaprakash B~S, Saulson P~R, Savage R, Sawadsky A,
  Scheuer J, Schilling R, Schmidt P, Schnabel R, Schofield R~M~S, Schreiber E,
  Schuette D, Schutz B~F, Scott J, Scott S~M, Sellers D, Sengupta A~S, Sergeev
  A, Serna G, Sevigny A, Shaddock D~A, Shahriar M~S, Shaltev M, Shao Z, Shapiro
  B, Shawhan P, Shoemaker D~H, Sidery T~L, Siemens X, Sigg D, Silva A~D,
  Simakov D, Singer A, Singer L, Singh R, Sintes A~M, Slagmolen B~J~J, Smith
  J~R, Smith M~R, Smith R~J~E, Smith-Lefebvre N~D, Son E~J, Sorazu B, Souradeep
  T, Staley A, Stebbins J, Steinke M, Steinlechner J, Steinlechner S,
  Steinmeyer D, Stephens B~C, Steplewski S, Stevenson S, Stone R, Strain K~A,
  Strigin S, Sturani R, Stuver A~L, Summerscales T~Z, Sutton P~J, Szczepanczyk
  M, Szeifert G, Talukder D, Tanner D~B, T{\'{a}}pai M, Tarabrin S~P,
  Taracchini A, Taylor R, Tellez G, Theeg T, Thirugnanasambandam M~P, Thomas M,
  Thomas P, Thorne K~A, Thorne K~S, Thrane E, Tiwari V, Tomlinson C, Torres
  C~V, Torrie C~I, Traylor G, Tse M, Tshilumba D, Ugolini D, Unnikrishnan C~S,
  Urban A~L, Usman S~A, Vahlbruch H, Vajente G, Valdes G, Vallisneri M, van
  Veggel A~A, Vass S, Vaulin R, Vecchio A, Veitch J, Veitch P~J, Venkateswara
  K, Vincent-Finley R, Vitale S, Vo T, Vorvick C, Vousden W~D, Vyatchanin S~P,
  Wade A~R, Wade L, Wade M, Walker M, Wallace L, Walsh S, Wang H, Wang M, Wang
  X, Ward R~L, Warner J, Was M, Weaver B, Weinert M, Weinstein A~J, Weiss R,
  Welborn T, Wen L, Wessels P, Westphal T, Wette K, Whelan J~T, Whitcomb S~E,
  White D~J, Whiting B~F, Wilkinson C, Williams L, Williams R, Williamson A~R,
  Willis J~L, Willke B, Wimmer M, Winkler W, Wipf C~C, Wittel H, Woan G, Worden
  J, Xie S, Yablon J, Yakushin I, Yam W, Yamamoto H, Yancey C~C, Yang Q,
  Zanolin M, Zhang F, Zhang L, Zhang M, Zhang Y, Zhao C, Zhou M, Zhu X~J,
  Zucker M~E, Zuraw S and Zweizig J 2015 {\em Classical and Quantum Gravity\/}
  {\bf 32} 074001
  \urlprefix\url{https://doi.org/10.1088\%2F0264-9381\%2F32\%2F7\%2F074001}

\bibitem{AcEA2015}
Acernese F, Agathos M, Agatsuma K, Aisa D, Allemandou N, Allocca A, Amarni J,
  Astone P, Balestri G, Ballardin G, Barone F, Baronick J~P, Barsuglia M, Basti
  A, Basti F, Bauer T~S, Bavigadda V, Bejger M, Beker M~G, Belczynski C,
  Bersanetti D, Bertolini A, Bitossi M, Bizouard M~A, Bloemen S, Blom M, Boer
  M, Bogaert G, Bondi D, Bondu F, Bonelli L, Bonnand R, Boschi V, Bosi L,
  Bouedo T, Bradaschia C, Branchesi M, Briant T, Brillet A, Brisson V, Bulik T,
  Bulten H~J, Buskulic D, Buy C, Cagnoli G, Calloni E, Campeggi C, Canuel B,
  Carbognani F, Cavalier F, Cavalieri R, Cella G, Cesarini E, Chassande-Mottin
  E, Chincarini A, Chiummo A, Chua S, Cleva F, Coccia E, Cohadon P~F, Colla A,
  Colombini M, Conte A, Coulon J~P, Cuoco E, Dalmaz A, D'Antonio S, Dattilo V,
  Davier M, Day R, Debreczeni G, Degallaix J, Del{\'{e}}glise S, Pozzo W~D,
  Dereli H, Rosa R~D, Fiore L~D, Lieto A~D, Virgilio A~D, Doets M, Dolique V,
  Drago M, Ducrot M, Endr{\H{o}}czi G, Fafone V, Farinon S, Ferrante I, Ferrini
  F, Fidecaro F, Fiori I, Flaminio R, Fournier J~D, Franco S, Frasca S,
  Frasconi F, Gammaitoni L, Garufi F, Gaspard M, Gatto A, Gemme G, Gendre B,
  Genin E, Gennai A, Ghosh S, Giacobone L, Giazotto A, Gouaty R, Granata M,
  Greco G, Groot P, Guidi G~M, Harms J, Heidmann A, Heitmann H, Hello P,
  Hemming G, Hennes E, Hofman D, Jaranowski P, Jonker R~J~G, Kasprzack M,
  K{\'{e}}f{\'{e}}lian F, Kowalska I, Kraan M, Kr{\'{o}}lak A, Kutynia A,
  Lazzaro C, Leonardi M, Leroy N, Letendre N, Li T~G~F, Lieunard B, Lorenzini
  M, Loriette V, Losurdo G, Magazz{\`{u}} C, Majorana E, Maksimovic I, Malvezzi
  V, Man N, Mangano V, Mantovani M, Marchesoni F, Marion F, Marque J, Martelli
  F, Martellini L, Masserot A, Meacher D, Meidam J, Mezzani F, Michel C, Milano
  L, Minenkov Y, Moggi A, Mohan M, Montani M, Morgado N, Mours B, Mul F, Nagy
  M~F, Nardecchia I, Naticchioni L, Nelemans G, Neri I, Neri M, Nocera F,
  Pacaud E, Palomba C, Paoletti F, Paoli A, Pasqualetti A, Passaquieti R,
  Passuello D, Perciballi M, Petit S, Pichot M, Piergiovanni F, Pillant G,
  Piluso A, Pinard L, Poggiani R, Prijatelj M, Prodi G~A, Punturo M, Puppo P,
  Rabeling D~S, R{\'{a}}cz I, Rapagnani P, Razzano M, Re V, Regimbau T, Ricci
  F, Robinet F, Rocchi A, Rolland L, Romano R, Rosi{\'{n}}ska D, Ruggi P,
  Saracco E, Sassolas B, Schimmel F, Sentenac D, Sequino V, Shah S, Siellez K,
  Straniero N, Swinkels B, Tacca M, Tonelli M, Travasso F, Turconi M, Vajente
  G, van Bakel N, van Beuzekom M, van~den Brand J~F~J, Broeck C~V~D, van~der
  Sluys M~V, van Heijningen J, Vas{\'{u}}th M, Vedovato G, Veitch J, Verkindt
  D, Vetrano F, Vicer{\'{e}} A, Vinet J~Y, Visser G, Vocca H, Ward R, Was M,
  Wei L~W, Yvert M, {\.{z}}ny A~Z and Zendri J~P 2014 {\em Classical and
  Quantum Gravity\/} {\bf 32} 024001
  \urlprefix\url{https://doi.org/10.1088%2F0264-9381%2F32%2F2%2F024001}

\bibitem{AbEA2019}
Abbott B~P, Abbott R, Abbott T~D, Abraham S, Acernese F, Ackley K, Adams C,
  Adhikari R~X, Adya V~B, Affeldt C, Agathos M, Agatsuma K, Aggarwal N, Aguiar
  O~D, Aiello L, Ain A, Ajith P, Allen G, Allocca A, Aloy M~A, Altin P~A, Amato
  A, Ananyeva A, Anderson S~B, Anderson W~G, Angelova S~V, Antier S, Appert S,
  Arai K, Araya M~C, Areeda J~S, Ar{\`{e}}ne M, Arnaud N, Arun K~G, Ascenzi S,
  Ashton G, Aston S~M, Astone P, Aubin F, Aufmuth P, AultONeal K, Austin C,
  Avendano V, Avila-Alvarez A, Babak S, Bacon P, Badaracco F, Bader M~K~M, Bae
  S, Baker P~T, Baldaccini F, Ballardin G, Ballmer S~W, Banagiri S, Barayoga
  J~C, Barclay S~E, Barish B~C, Barker D, Barkett K, Barnum S, Barone F, Barr
  B, Barsotti L, Barsuglia M, Barta D, Bartlett J, Bartos I, Bassiri R, Basti
  A, Bawaj M, Bayley J~C, Bazzan M, B{\'{e}}csy B, Bejger M, Belahcene I, Bell
  A~S, Beniwal D, Berger B~K, Bergmann G, Bernuzzi S, Bero J~J, Berry C~P~L,
  Bersanetti D, Bertolini A, Betzwieser J, Bhandare R, Bidler J, Bilenko I~A,
  Bilgili S~A, Billingsley G, Birch J, Birney R, Birnholtz O, Biscans S,
  Biscoveanu S, Bisht A, Bitossi M, Bizouard M~A, Blackburn J~K, Blair C~D,
  Blair D~G, Blair R~M, Bloemen S, Bode N, Boer M, Boetzel Y, Bogaert G, Bondu
  F, Bonilla E, Bonnand R, Booker P, Boom B~A, Booth C~D, Bork R, Boschi V,
  Bose S, Bossie K, Bossilkov V, Bosveld J, Bouffanais Y, Bozzi A, Bradaschia
  C, Brady P~R, Bramley A, Branchesi M, Brau J~E, Briant T, Briggs J~H,
  Brighenti F, Brillet A, Brinkmann M, Brisson V, Brockill P, Brooks A~F, Brown
  D~D, Brunett S, Buikema A, Bulik T, Bulten H~J, Buonanno A, Buscicchio R,
  Buskulic D, Buy C, Byer R~L, Cabero M, Cadonati L, Cagnoli G, Cahillane C,
  Bustillo J~C, Callister T~A, Calloni E, Camp J~B, Campbell W~A, Canepa M,
  Cannon K~C, Cao H, Cao J, Capocasa E, Carbognani F, Caride S, Carney M~F,
  Carullo G, Diaz J~C, Casentini C, Caudill S, Cavagli{\`{a}} M, Cavalier F,
  Cavalieri R, Cella G, Cerd{\'{a}}-Dur{\'{a}}n P, Cerretani G, Cesarini E,
  Chaibi O, Chakravarti K, Chamberlin S~J, Chan M, Chao S, Charlton P, Chase
  E~A, Chassande-Mottin E, Chatterjee D, Chaturvedi M, Chatziioannou K,
  Cheeseboro B~D, Chen H~Y, Chen X, Chen Y, Cheng H~P, Cheong C~K, Chia H~Y,
  Chincarini A, Chiummo A, Cho G, Cho H~S, Cho M, Christensen N, Chu Q, Chua S,
  Chung K~W, Chung S, Ciani G, Ciobanu A~A, Ciolfi R, Cipriano F, Cirone A,
  Clara F, Clark J~A, Clearwater P, Cleva F, Cocchieri C, Coccia E, Cohadon
  P~F, Cohen D, Colgan R, Colleoni M, Collette C~G, Collins C, Cominsky L~R, Jr
  M~C, Conti L, Cooper S~J, Corban P, Corbitt T~R, Cordero-Carri{\'{o}}n I,
  Corley K~R, Cornish N, Corsi A, Cortese S, Costa C~A, Cotesta R, Coughlin
  M~W, Coughlin S~B, Coulon J~P, Countryman S~T, Couvares P, Covas P~B, Cowan
  E~E, Coward D~M, Cowart M~J, Coyne D~C, Coyne R, Creighton J~D~E, Creighton
  T~D, Cripe J, Croquette M, Crowder S~G, Cullen T~J, Cumming A, Cunningham L,
  Cuoco E, Canton T~D, D{\'{a}}lya G, Danilishin S~L, D'Antonio S, Danzmann K,
  Dasgupta A, Costa C~F~D~S, Datrier L~E~H, Dattilo V, Dave I, Davier M, Davis
  D, Daw E~J, DeBra D, Deenadayalan M, Degallaix J, Laurentis M~D,
  Del{\'{e}}glise S, Pozzo W~D, DeMarchi L~M, Demos N, Dent T, Pietri R~D,
  Derby J, Rosa R~D, Rossi C~D, DeSalvo R, de~Varona O, Dhurandhar S,
  D{\'{\i}}az M~C, Dietrich T, Fiore L~D, Giovanni M~D, Girolamo T~D, Lieto
  A~D, Ding B, Pace S~D, Palma I~D, Renzo F~D, Dmitriev A, Doctor Z, Donovan F,
  Dooley K~L, Doravari S, Dorrington I, Downes T~P, Drago M, Driggers J~C, Du
  Z, Ducoin J~G, Dupej P, Dwyer S~E, Easter P~J, Edo T~B, Edwards M~C, Effler
  A, Ehrens P, Eichholz J, Eikenberry S~S, Eisenmann M, Eisenstein R~A, Essick
  R~C, Estelles H, Estevez D, Etienne Z~B, Etzel T, Evans M, Evans T~M, Fafone
  V, Fair H, Fairhurst S, Fan X, Farinon S, Farr B, Farr W~M, Fauchon-Jones
  E~J, Favata M, Fays M, Fazio M, Fee C, Feicht J, Fejer M~M, Feng F,
  Fernandez-Galiana A, Ferrante I, Ferreira E~C, Ferreira T~A, Ferrini F,
  Fidecaro F, Fiori I, Fiorucci D, Fishbach M, Fisher R~P, Fishner J~M,
  Fitz-Axen M, Flaminio R, Fletcher M, Flynn E, Fong H, Font J~A, Forsyth
  P~W~F, Fournier J~D, Frasca S, Frasconi F, Frei Z, Freise A, Frey R, Frey V,
  Fritschel P, Frolov V~V, Fulda P, Fyffe M, Gabbard H~A, Gadre B~U, Gaebel
  S~M, Gair J~R, Gammaitoni L, Ganija M~R, Gaonkar S~G, Garcia A,
  Garc{\'{\i}}a-Quir{\'{o}}s C, Garufi F, Gateley B, Gaudio S, Gaur G, Gayathri
  V, Gemme G, Genin E, Gennai A, George D, George J, Gergely L, Germain V,
  Ghonge S, Ghosh A, Ghosh A, Ghosh S, Giacomazzo B, Giaime J~A, Giardina K~D,
  Giazotto A, Gill K, Giordano G, Glover L, Godwin P, Goetz E, Goetz R,
  Goncharov B, Gonz{\'{a}}lez G, Castro J~M~G, Gopakumar A, Gorodetsky M~L,
  Gossan S~E, Gosselin M, Gouaty R, Grado A, Graef C, Granata M, Grant A, Gras
  S, Grassia P, Gray C, Gray R, Greco G, Green A~C, Green R, Gretarsson E~M,
  Groot P, Grote H, Grunewald S, Gruning P, Guidi G~M, Gulati H~K, Guo Y, Gupta
  A, Gupta M~K, Gustafson E~K, Gustafson R, Haegel L, Halim O, Hall B~R, Hall
  E~D, Hamilton E~Z, Hammond G, Haney M, Hanke M~M, Hanks J, Hanna C, Hannam
  M~D, Hannuksela O~A, Hanson J, Hardwick T, Haris K, Harms J, Harry G~M, Harry
  I~W, Haster C~J, Haughian K, Hayes F~J, Healy J, Heidmann A, Heintze M~C,
  Heitmann H, Hello P, Hemming G, Hendry M, Heng I~S, Hennig J, Heptonstall
  A~W, Vivanco F~H, Heurs M, Hild S, Hinderer T, Hoak D, Hochheim S, Hofman D,
  Holgado A~M, Holland N~A, Holt K, Holz D~E, Hopkins P, Horst C, Hough J,
  Howell E~J, Hoy C~G, Hreibi A, Huerta E~A, Huet D, Hughey B, Hulko M, Husa S,
  Huttner S~H, Huynh-Dinh T, Idzkowski B, Iess A, Ingram C, Inta R, Intini G,
  Irwin B, Isa H~N, Isac J~M, Isi M, Iyer B~R, Izumi K, Jacqmin T, Jadhav S~J,
  Jani K, Janthalur N~N, Jaranowski P, Jenkins A~C, Jiang J, Johnson D~S, Jones
  A~W, Jones D~I, Jones R, Jonker R~J~G, Ju L, Junker J, Kalaghatgi C~V,
  Kalogera V, Kamai B, Kandhasamy S, Kang G, Kanner J~B, Kapadia S~J, Karki S,
  Karvinen K~S, Kashyap R, Kasprzack M, Katsanevas S, Katsavounidis E, Katzman
  W, Kaufer S, Kawabe K, Keerthana N~V, K{\'{e}}f{\'{e}}lian F, Keitel D,
  Kennedy R, Key J~S, Khalili F~Y, Khan H, Khan I, Khan S, Khan Z, Khazanov
  E~A, Khursheed M, Kijbunchoo N, Kim C, Kim J~C, Kim K, Kim W, Kim W~S, Kim
  Y~M, Kimball C, King E~J, King P~J, Kinley-Hanlon M, Kirchhoff R, Kissel J~S,
  Kleybolte L, Klika J~H, Klimenko S, Knowles T~D, Koch P, Koehlenbeck S~M,
  Koekoek G, Koley S, Kondrashov V, Kontos A, Koper N, Korobko M, Korth W~Z,
  Kowalska I, Kozak D~B, Kringel V, Krishnendu N, Kr{\'{o}}lak A, Kuehn G,
  Kumar A, Kumar P, Kumar R, Kumar S, Kuo L, Kutynia A, Kwang S, Lackey B~D,
  Lai K~H, Lam T~L, Landry M, Lane B~B, Lang R~N, Lange J, Lantz B, Lanza R~K,
  Lartaux-Vollard A, Lasky P~D, Laxen M, Lazzarini A, Lazzaro C, Leaci P,
  Leavey S, Lecoeuche Y~K, Lee C~H, Lee H~K, Lee H~M, Lee H~W, Lee J, Lee K,
  Lehmann J, Lenon A, Leroy N, Letendre N, Levin Y, Li J, Li K~J~L, Li T~G~F,
  Li X, Lin F, Linde F, Linker S~D, Littenberg T~B, Liu J, Liu X, Lo R~K~L,
  Lockerbie N~A, London L~T, Longo A, Lorenzini M, Loriette V, Lormand M,
  Losurdo G, Lough J~D, Lousto C~O, Lovelace G, Lower M~E, Lück H, Lumaca D,
  Lundgren A~P, Lynch R, Ma Y, Macas R, Macfoy S, MacInnis M, Macleod D~M,
  Macquet A, Maga{\~{n}}a-Sandoval F, Zertuche L~M, Magee R~M, Majorana E,
  Maksimovic I, Malik A, Man N, Mandic V, Mangano V, Mansell G~L, Manske M,
  Mantovani M, Mapelli M, Marchesoni F, Marion F, M{\'{a}}rka S, M{\'{a}}rka Z,
  Markakis C, Markosyan A~S, Markowitz A, Maros E, Marquina A, Marsat S,
  Martelli F, Martin I~W, Martin R~M, Martynov D~V, Mason K, Massera E,
  Masserot A, Massinger T~J, Masso-Reid M, Mastrogiovanni S, Matas A, Matichard
  F, Matone L, Mavalvala N, Mazumder N, McCann J~J, McCarthy R, McClelland D~E,
  McCormick S, McCuller L, McGuire S~C, McIver J, McManus D~J, McRae T,
  McWilliams S~T, Meacher D, Meadors G~D, Mehmet M, Mehta A~K, Meidam J,
  Melatos A, Mendell G, Mercer R~A, Mereni L, Merilh E~L, Merzougui M, Meshkov
  S, Messenger C, Messick C, Metzdorff R, Meyers P~M, Miao H, Michel C,
  Middleton H, Mikhailov E~E, Milano L, Miller A~L, Miller A, Millhouse M,
  Mills J~C, Milovich-Goff M~C, Minazzoli O, Minenkov Y, Mishkin A, Mishra C,
  Mistry T, Mitra S, Mitrofanov V~P, Mitselmakher G, Mittleman R, Mo G, Moffa
  D, Mogushi K, Mohapatra S~R~P, Montani M, Moore C~J, Moraru D, Moreno G,
  Morisaki S, Mours B, Mow-Lowry C~M, Mukherjee A, Mukherjee D, Mukherjee S,
  Mukund N, Mullavey A, Munch J, Mu{\~{n}}iz E~A, Muratore M, Murray P~G, Nagar
  A, Nardecchia I, Naticchioni L, Nayak R~K, Neilson J, Nelemans G, Nelson
  T~J~N, Nery M, Neunzert A, Ng K~Y, Ng S, Nguyen P, Nichols D, Nissanke S,
  Nocera F, North C, Nuttall L~K, Obergaulinger M, Oberling J, O'Brien B~D,
  O'Dea G~D, Ogin G~H, Oh J~J, Oh S~H, Ohme F, Ohta H, Okada M~A, Oliver M,
  Oppermann P, Oram R~J, O'Reilly B, Ormiston R~G, Ortega L~F, O'Shaughnessy R,
  Ossokine S, Ottaway D~J, Overmier H, Owen B~J, Pace A~E, Pagano G, Page M~A,
  Pai A, Pai S~A, Palamos J~R, Palashov O, Palomba C, Pal-Singh A, Pan H~W,
  Pang B, Pang P~T~H, Pankow C, Pannarale F, Pant B~C, Paoletti F, Paoli A,
  Parida A, Parker W, Pascucci D, Pasqualetti A, Passaquieti R, Passuello D,
  Patil M, Patricelli B, Pearlstone B~L, Pedersen C, Pedraza M, Pedurand R,
  Pele A, Penn S, Perez C~J, Perreca A, Pfeiffer H~P, Phelps M, Phukon K~S,
  Piccinni O~J, Pichot M, Piergiovanni F, Pillant G, Pinard L, Pirello M,
  Pitkin M, Poggiani R, Pong D~Y~T, Ponrathnam S, Popolizio P, Porter E~K,
  Powell J, Prajapati A~K, Prasad J, Prasai K, Prasanna R, Pratten G,
  Prestegard T, Privitera S, Prodi G~A, Prokhorov L~G, Puncken O, Punturo M,
  Puppo P, Pürrer M, Qi H, Quetschke V, Quinonez P~J, Quintero E~A,
  Quitzow-James R, Raab F~J, Radkins H, Radulescu N, Raffai P, Raja S, Rajan C,
  Rajbhandari B, Rakhmanov M, Ramirez K~E, Ramos-Buades A, Rana J, Rao K,
  Rapagnani P, Raymond V, Razzano M, Read J, Regimbau T, Rei L, Reid S, Reitze
  D~H, Ren W, Ricci F, Richardson C~J, Richardson J~W, Ricker P~M, Riles K,
  Rizzo M, Robertson N~A, Robie R, Robinet F, Rocchi A, Rolland L, Rollins J~G,
  Roma V~J, Romanelli M, Romano R, Romel C~L, Romie J~H, Rose K, Rosi{\'{n}}ska
  D, Rosofsky S~G, Ross M~P, Rowan S, Rüdiger A, Ruggi P, Rutins G, Ryan K,
  Sachdev S, Sadecki T, Sakellariadou M, Salconi L, Saleem M, Samajdar A,
  Sammut L, Sanchez E~J, Sanchez L~E, Sanchis-Gual N, Sandberg V, Sanders J~R,
  Santiago K~A, Sarin N, Sassolas B, Sathyaprakash B~S, Saulson P~R, Sauter O,
  Savage R~L, Schale P, Scheel M, Scheuer J, Schmidt P, Schnabel R, Schofield
  R~M~S, Schönbeck A, Schreiber E, Schulte B~W, Schutz B~F, Schwalbe S~G,
  Scott J, Scott S~M, Seidel E, Sellers D, Sengupta A~S, Sennett N, Sentenac D,
  Sequino V, Sergeev A, Setyawati Y, Shaddock D~A, Shaffer T, Shahriar M~S,
  Shaner M~B, Shao L, Sharma P, Shawhan P, Shen H, Shink R, Shoemaker D~H,
  Shoemaker D~M, ShyamSundar S, Siellez K, Sieniawska M, Sigg D, Silva A~D,
  Singer L~P, Singh N, Singhal A, Sintes A~M, Sitmukhambetov S, Skliris V,
  Slagmolen B~J~J, Slaven-Blair T~J, Smith J~R, Smith R~J~E, Somala S, Son E~J,
  Sorazu B, Sorrentino F, Souradeep T, Sowell E, Spencer A~P, Spera M,
  Srivastava A~K, Srivastava V, Staats K, Stachie C, Standke M, Steer D~A,
  Steinke M, Steinlechner J, Steinlechner S, Steinmeyer D, Stevenson S~P,
  Stocks D, Stone R, Stops D~J, Strain K~A, Stratta G, Strigin S~E, Strunk A,
  Sturani R, Stuver A~L, Sudhir V, Summerscales T~Z, Sun L, Sunil S, Suresh J,
  Sutton P~J, Swinkels B~L, Szczepa{\'{n}}czyk M~J, Tacca M, Tait S~C, Talbot
  C, Talukder D, Tanner D~B, T{\'{a}}pai M, Taracchini A, Tasson J~D, Taylor R,
  Thies F, Thomas M, Thomas P, Thondapu S~R, Thorne K~A, Thrane E, Tiwari S,
  Tiwari S, Tiwari V, Toland K, Tonelli M, Tornasi Z, Torres-Forn{\'{e}} A,
  Torrie C~I, Töyrä D, Travasso F, Traylor G, Tringali M~C, Trovato A, Trozzo
  L, Trudeau R, Tsang K~W, Tse M, Tso R, Tsukada L, Tsuna D, Tuyenbayev D, Ueno
  K, Ugolini D, Unnikrishnan C~S, Urban A~L, Usman S~A, Vahlbruch H, Vajente G,
  Valdes G, van Bakel N, van Beuzekom M, van~den Brand J~F~J, Broeck C~V~D,
  Vander-Hyde D~C, van~der Schaaf L, van Heijningen J~V, van Veggel A~A,
  Vardaro M, Varma V, Vass S, Vas{\'{u}}th M, Vecchio A, Vedovato G, Veitch J,
  Veitch P~J, Venkateswara K, Venugopalan G, Verkindt D, Vetrano F,
  Vicer{\'{e}} A, Viets A~D, Vine D~J, Vinet J~Y, Vitale S, Vo T, Vocca H,
  Vorvick C, Vyatchanin S~P, Wade A~R, Wade L~E, Wade M, Walet R, Walker M,
  Wallace L, Walsh S, Wang G, Wang H, Wang J~Z, Wang W~H, Wang Y~F, Ward R~L,
  Warden Z~A, Warner J, Was M, Watchi J, Weaver B, Wei L~W, Weinert M,
  Weinstein A~J, Weiss R, Wellmann F, Wen L, Wessel E~K, We{\ss}els P,
  Westhouse J~W, Wette K, Whelan J~T, Whiting B~F, Whittle C, Wilken D~M,
  Williams D, Williamson A~R, Willis J~L, Willke B, Wimmer M~H, Winkler W, Wipf
  C~C, Wittel H, Woan G, Woehler J, Wofford J~K, Worden J, Wright J~L, Wu D~S,
  Wysocki D~M, Xiao L, Yamamoto H, Yancey C~C, Yang L, Yap M~J, Yazback M,
  Yeeles D~W, Yu H, Yu H, Yuen S~H~R, Yvert M, Zadro{\.{z}}ny A~K, Zanolin M,
  Zelenova T, Zendri J~P, Zevin M, Zhang J, Zhang L, Zhang T, Zhao C, Zhou M,
  Zhou Z, Zhu X~J, Zimmerman A~B, Zlochower Y, Zucker M~E and and J~Z 2019 {\em
  The Astrophysical Journal\/} {\bf 882} L24
  \urlprefix\url{https://doi.org/10.3847%2F2041-8213%2Fab3800}

\bibitem{AbEA2021}
Abbott R, Abbott T~D, Abraham S, Acernese F, Ackley K, Adams A, Adams C,
  Adhikari R~X, Adya V~B, Affeldt C, Agathos M, Agatsuma K, Aggarwal N, Aguiar
  O~D, Aiello L, Ain A, Ajith P, Akcay S, Allen G, Allocca A, Altin P~A, Amato
  A, Anand S, Ananyeva A, Anderson S~B, Anderson W~G, Angelova S~V, Ansoldi S,
  Antelis J~M, Antier S, Appert S, Arai K, Araya M~C, Areeda J~S, Ar\`ene M,
  Arnaud N, Aronson S~M, Arun K~G, Asali Y, Ascenzi S, Ashton G, Aston S~M,
  Astone P, Aubin F, Aufmuth P, AultONeal K, Austin C, Avendano V, Babak S,
  Badaracco F, Bader M~K~M, Bae S, Baer A~M, Bagnasco S, Baird J, Ball M,
  Ballardin G, Ballmer S~W, Bals A, Balsamo A, Baltus G, Banagiri S, Bankar D,
  Bankar R~S, Barayoga J~C, Barbieri C, Barish B~C, Barker D, Barneo P, Barnum
  S, Barone F, Barr B, Barsotti L, Barsuglia M, Barta D, Bartlett J, Bartos I,
  Bassiri R, Basti A, Bawaj M, Bayley J~C, Bazzan M, Becher B~R, B\'ecsy B,
  Bedakihale V~M, Bejger M, Belahcene I, Beniwal D, Benjamin M~G, Bennett T~F,
  Bentley J~D, Bergamin F, Berger B~K, Bergmann G, Bernuzzi S, Berry C~P~L,
  Bersanetti D, Bertolini A, Betzwieser J, Bhandare R, Bhandari A~V,
  Bhattacharjee D, Bidler J, Bilenko I~A, Billingsley G, Birney R, Birnholtz O,
  Biscans S, Bischi M, Biscoveanu S, Bisht A, Bitossi M, Bizouard M~A,
  Blackburn J~K, Blackman J, Blair C~D, Blair D~G, Blair R~M, Blanch O, Bobba
  F, Bode N, Boer M, Boetzel Y, Bogaert G, Boldrini M, Bondu F, Bonilla E,
  Bonnand R, Booker P, Boom B~A, Bork R, Boschi V, Bose S, Bossilkov V, Boudart
  V, Bouffanais Y, Bozzi A, Bradaschia C, Brady P~R, Bramley A, Branchesi M,
  Brau J~E, Breschi M, Briant T, Briggs J~H, Brighenti F, Brillet A, Brinkmann
  M, Brockill P, Brooks A~F, Brooks J, Brown D~D, Brunett S, Bruno G, Bruntz R,
  Buikema A, Bulik T, Bulten H~J, Buonanno A, Buscicchio R, Buskulic D, Byer
  R~L, Cabero M, Cadonati L, Caesar M, Cagnoli G, Cahillane C,
  Calder\'on~Bustillo J, Callaghan J~D, Callister T~A, Calloni E, Camp J~B,
  Canepa M, Cannon K~C, Cao H, Cao J, Carapella G, Carbognani F, Carney M~F,
  Carpinelli M, Carullo G, Carver T~L, Casanueva~Diaz J, Casentini C, Caudill
  S, Cavagli\`a M, Cavalier F, Cavalieri R, Cella G, Cerd\'a-Dur\'an P,
  Cesarini E, Chaibi W, Chakravarti K, Chan C~L, Chan C, Chandra K, Chanial P,
  Chao S, Charlton P, Chase E~A, Chassande-Mottin E, Chatterjee D,
  Chattopadhyay D, Chaturvedi M, Chatziioannou K, Chen A, Chen H~Y, Chen X,
  Chen Y, Cheng H~P, Cheong C~K, Chia H~Y, Chiadini F, Chierici R, Chincarini
  A, Chiummo A, Cho G, Cho H~S, Cho M, Choate S, Christensen N, Chu Q, Chua S,
  Chung K~W, Chung S, Ciani G, Ciecielag P, Cie\ifmmode~\acute{s}\else
  \'{s}\fi{}lar M, Cifaldi M, Ciobanu A~A, Ciolfi R, Cipriano F, Cirone A,
  Clara F, Clark E~N, Clark J~A, Clarke L, Clearwater P, Clesse S, Cleva F,
  Coccia E, Cohadon P~F, Cohen D~E, Colleoni M, Collette C~G, Collins C, Colpi
  M, Constancio M, Conti L, Cooper S~J, Corban P, Corbitt T~R,
  Cordero-Carri\'on I, Corezzi S, Corley K~R, Cornish N, Corre D, Corsi A,
  Cortese S, Costa C~A, Cotesta R, Coughlin M~W, Coughlin S~B, Coulon J~P,
  Countryman S~T, Cousins B, Couvares P, Covas P~B, Coward D~M, Cowart M~J,
  Coyne D~C, Coyne R, Creighton J~D~E, Creighton T~D, Croquette M, Crowder S~G,
  Cudell J~R, Cullen T~J, Cumming A, Cummings R, Cunningham L, Cuoco E,
  Cury\l{}o M, Canton T~D, D\'alya G, Dana A, DaneshgaranBajastani L~M,
  D'Angelo B, Danila B, Danilishin S~L, D'Antonio S, Danzmann K, Darsow-Fromm
  C, Dasgupta A, Datrier L~E~H, Dattilo V, Dave I, Davier M, Davies G~S, Davis
  D, Daw E~J, Dean R, DeBra D, Deenadayalan M, Degallaix J, De~Laurentis M,
  Del\'eglise S, Del~Favero V, De~Lillo F, De~Lillo N, Del~Pozzo W, DeMarchi
  L~M, De~Matteis F, D'Emilio V, Demos N, Denker T, Dent T, Depasse A,
  De~Pietri R, De~Rosa R, De~Rossi C, DeSalvo R, de~Varona O, Dhurandhar S,
  D\'{\i}az M~C, Diaz-Ortiz M, Didio N~A, Dietrich T, Di~Fiore L, DiFronzo C,
  Di~Giorgio C, Di~Giovanni F, Di~Giovanni M, Di~Girolamo T, Di~Lieto A, Ding
  B, Di~Pace S, Di~Palma I, Di~Renzo F, Divakarla A~K, Dmitriev A, Doctor Z,
  D'Onofrio L, Donovan F, Dooley K~L, Doravari S, Dorrington I, Downes T~P,
  Drago M, Driggers J~C, Du Z, Ducoin J~G, Dupej P, Durante O, D'Urso D,
  Duverne P~A, Dwyer S~E, Easter P~J, Eddolls G, Edelman B, Edo T~B, Edy O,
  Effler A, Eichholz J, Eikenberry S~S, Eisenmann M, Eisenstein R~A, Ejlli A,
  Errico L, Essick R~C, Estell\'es H, Estevez D, Etienne Z~B, Etzel T, Evans M,
  Evans T~M, Ewing B~E, Fafone V, Fair H, Fairhurst S, Fan X, Farah A~M,
  Farinon S, Farr B, Farr W~M, Fauchon-Jones E~J, Favata M, Fays M, Fazio M,
  Feicht J, Fejer M~M, Feng F, Fenyvesi E, Ferguson D~L, Fernandez-Galiana A,
  Ferrante I, Ferreira T~A, Fidecaro F, Figura P, Fiori I, Fiorucci D, Fishbach
  M, Fisher R~P, Fishner J~M, Fittipaldi R, Fitz-Axen M, Fiumara V, Flaminio R,
  Floden E, Flynn E, Fong H, Font J~A, Forsyth P~W~F, Fournier J~D, Frasca S,
  Frasconi F, Frei Z, Freise A, Frey R, Frey V, Fritschel P, Frolov V~V,
  Fronz\'e G~G, Fulda P, Fyffe M, Gabbard H~A, Gadre B~U, Gaebel S~M, Gair J~R,
  Gais J, Galaudage S, Gamba R, Ganapathy D, Ganguly A, Gaonkar S~G, Garaventa
  B, Garc\'{\i}a-Quir\'os C, Garufi F, Gateley B, Gaudio S, Gayathri V, Gemme
  G, Gennai A, George D, George J, George R~N, Gergely L, Ghonge S, Ghosh A,
  Ghosh A, Ghosh S, Giacomazzo B, Giacoppo L, Giaime J~A, Giardina K~D, Gibson
  D~R, Gier C, Gill K, Giri P, Glanzer J, Gleckl A~E, Godwin P, Goetz E, Goetz
  R, Gohlke N, Goncharov B, Gonz\'alez G, Gopakumar A, Gossan S~E, Gosselin M,
  Gouaty R, Grace B, Grado A, Granata M, Granata V, Grant A, Gras S, Grassia P,
  Gray C, Gray R, Greco G, Green A~C, Green R, Gretarsson E~M, Griggs H~L,
  Grignani G, Grimaldi A, Grimes E, Grimm S~J, Grote H, Grunewald S, Gruning P,
  Guerrero J~G, Guidi G~M, Guimaraes A~R, Guix\'e G, Gulati H~K, Guo Y, Gupta
  A, Gupta A, Gupta P, Gustafson E~K, Gustafson R, Guzman F, Haegel L, Halim O,
  Hall E~D, Hamilton E~Z, Hammond G, Haney M, Hanke M~M, Hanks J, Hanna C,
  Hannam M~D, Hannuksela O~A, Hannuksela O, Hansen H, Hansen T~J, Hanson J,
  Harder T, Hardwick T, Haris K, Harms J, Harry G~M, Harry I~W, Hartwig D,
  Hasskew R~K, Haster C~J, Haughian K, Hayes F~J, Healy J, Heidmann A, Heintze
  M~C, Heinze J, Heinzel J, Heitmann H, Hellman F, Hello P, Helmling-Cornell
  A~F, Hemming G, Hendry M, Heng I~S, Hennes E, Hennig J, Hennig M~H,
  Hernandez~Vivanco F, Heurs M, Hild S, Hill P, Hines A~S, Hochheim S, Hofgard
  E, Hofman D, Hohmann J~N, Holgado A~M, Holland N~A, Hollows I~J, Holmes Z~J,
  Holt K, Holz D~E, Hopkins P, Horst C, Hough J, Howell E~J, Hoy C~G, Hoyland
  D, Huang Y, H\"ubner M~T, Huddart A~D, Huerta E~A, Hughey B, Hui V, Husa S,
  Huttner S~H, Hutzler B~M, Huxford R, Huynh-Dinh T, Idzkowski B, Iess A,
  Imperato S, Inchauspe H, Ingram C, Intini G, Isi M, Iyer B~R, JaberianHamedan
  V, Jacqmin T, Jadhav S~J, Jadhav S~P, James A~L, Jani K, Janssens K,
  Janthalur N~N, Jaranowski P, Jariwala D, Jaume R, Jenkins A~C, Jeunon M,
  Jiang J, Johns G~R, Johnson-McDaniel N~K, Jones A~W, Jones D~I, Jones J~D,
  Jones P, Jones R, Jonker R~J~G, Ju L, Junker J, Kalaghatgi C~V, Kalogera V,
  Kamai B, Kandhasamy S, Kang G, Kanner J~B, Kapadia S~J, Kapasi D~P,
  Karathanasis C, Karki S, Kashyap R, Kasprzack M, Kastaun W, Katsanevas S,
  Katsavounidis E, Katzman W, Kawabe K, K\'ef\'elian F, Keitel D, Key J~S,
  Khadka S, Khalili F~Y, Khan I, Khan S, Khazanov E~A, Khetan N, Khursheed M,
  Kijbunchoo N, Kim C, Kim G~J, Kim J~C, Kim K, Kim W~S, Kim Y~M, Kimball C,
  King P~J, Kinley-Hanlon M, Kirchhoff R, Kissel J~S, Kleybolte L, Klimenko S,
  Knowles T~D, Knyazev E, Koch P, Koehlenbeck S~M, Koekoek G, Koley S, Kolstein
  M, Komori K, Kondrashov V, Kontos A, Koper N, Korobko M, Korth W~Z, Kovalam
  M, Kozak D~B, Kr\"amer C, Kringel V, Krishnendu N~V, Kr\'olak A, Kuehn G,
  Kumar A, Kumar P, Kumar R, Kumar R, Kuns K, Kwang S, Lackey B~D, Laghi D,
  Lalande E, Lam T~L, Lamberts A, Landry M, Lane B~B, Lang R~N, Lange J, Lantz
  B, Lanza R~K, La~Rosa I, Lartaux-Vollard A, Lasky P~D, Laxen M, Lazzarini A,
  Lazzaro C, Leaci P, Leavey S, Lecoeuche Y~K, Lee H~M, Lee H~W, Lee J, Lee K,
  Lehmann J, Leon E, Leroy N, Letendre N, Levin Y, Li A, Li J, Li K~J~L, Li
  T~G~F, Li X, Linde F, Linker S~D, Linley J~N, Littenberg T~B, Liu J, Liu X,
  Llorens-Monteagudo M, Lo R~K~L, Lockwood A, London L~T, Longo A, Lorenzini M,
  Loriette V, Lormand M, Losurdo G, Lough J~D, Lousto C~O, Lovelace G, L\"uck
  H, Lumaca D, Lundgren A~P, Ma Y, Macas R, MacInnis M, Macleod D~M, MacMillan
  I~A~O, Macquet A, Maga\~na Hernandez I, Maga\~na Sandoval F, Magazz\`u C,
  Magee R~M, Majorana E, Maksimovic I, Maliakal S, Malik A, Man N, Mandic V,
  Mangano V, Mansell G~L, Manske M, Mantovani M, Mapelli M, Marchesoni F,
  Marion F, M\'arka S, M\'arka Z, Markakis C, Markosyan A~S, Markowitz A, Maros
  E, Marquina A, Marsat S, Martelli F, Martin I~W, Martin R~M, Martinez M,
  Martinez V, Martynov D~V, Masalehdan H, Mason K, Massera E, Masserot A,
  Massinger T~J, Masso-Reid M, Mastrogiovanni S, Matas A, Mateu-Lucena M,
  Matichard F, Matiushechkina M, Mavalvala N, Maynard E, McCann J~J, McCarthy
  R, McClelland D~E, McCormick S, McCuller L, McGuire S~C, McIsaac C, McIver J,
  McManus D~J, McRae T, McWilliams S~T, Meacher D, Meadors G~D, Mehmet M, Mehta
  A~K, Melatos A, Melchor D~A, Mendell G, Menendez-Vazquez A, Mercer R~A,
  Mereni L, Merfeld K, Merilh E~L, Merritt J~D, Merzougui M, Meshkov S,
  Messenger C, Messick C, Metzdorff R, Meyers P~M, Meylahn F, Mhaske A, Miani
  A, Miao H, Michaloliakos I, Michel C, Middleton H, Milano L, Miller A~L,
  Millhouse M, Mills J~C, Milotti E, Milovich-Goff M~C, Minazzoli O, Minenkov
  Y, Mir L~M, Mishkin A, Mishra C, Mistry T, Mitra S, Mitrofanov V~P,
  Mitselmakher G, Mittleman R, Mo G, Mogushi K, Mohapatra S~R~P, Mohite S~R,
  Molina I, Molina-Ruiz M, Mondin M, Montani M, Moore C~J, Moraru D, Morawski
  F, Moreno G, Morisaki S, Mours B, Mow-Lowry C~M, Mozzon S, Muciaccia F,
  Mukherjee A, Mukherjee D, Mukherjee S, Mukherjee S, Mukund N, Mullavey A,
  Munch J, Mu\~niz E~A, Murray P~G, Nadji S~L, Nagar A, Nardecchia I,
  Naticchioni L, Nayak R~K, Neil B~F, Neilson J, Nelemans G, Nelson T~J~N, Nery
  M, Neunzert A, Nitz A~H, Ng K~Y, Ng S, Nguyen C, Nguyen P, Nguyen T, Nichols
  S~A, Nissanke S, Nocera F, Noh M, North C, Nothard D, Nuttall L~K, Oberling
  J, O'Brien B~D, O'Dell J, Oganesyan G, Ogin G~H, Oh J~J, Oh S~H, Ohme F, Ohta
  H, Okada M~A, Olivetto C, Oppermann P, Oram R~J, O'Reilly B, Ormiston R~G,
  Ortega L~F, O'Shaughnessy R, Ossokine S, Osthelder C, Ottaway D~J, Overmier
  H, Owen B~J, Pace A~E, Pagano G, Page M~A, Pagliaroli G, Pai A, Pai S~A,
  Palamos J~R, Palashov O, Palomba C, Pan H, Panda P~K, Pang T~H, Pankow C,
  Pannarale F, Pant B~C, Paoletti F, Paoli A, Paolone A, Parker W, Pascucci D,
  Pasqualetti A, Passaquieti R, Passuello D, Patel M, Patricelli B, Payne E,
  Pechsiri T~C, Pedraza M, Pegoraro M, Pele A, Penn S, Perego A, Perez C~J,
  P\'erigois C, Perreca A, Perri\`es S, Petermann J, Petterson D, Pfeiffer H~P,
  Pham K~A, Phukon K~S, Piccinni O~J, Pichot M, Piendibene M, Piergiovanni F,
  Pierini L, Pierro V, Pillant G, Pilo F, Pinard L, Pinto I~M, Piotrzkowski K,
  Pirello M, Pitkin M, Placidi E, Plastino W, Pluchar C, Poggiani R, Polini E,
  Pong D~Y~T, Ponrathnam S, Popolizio P, Porter E~K, Poverman A, Powell J,
  Pracchia M, Prajapati A~K, Prasai K, Prasanna R, Pratten G, Prestegard T,
  Principe M, Prodi G~A, Prokhorov L, Prosposito P, Prudenzi L, Puecher A,
  Punturo M, Puosi F, Puppo P, P\"urrer M, Qi H, Quetschke V, Quinonez P~J,
  Quitzow-James R, Raab F~J, Raaijmakers G, Radkins H, Radulesco N, Raffai P,
  Rafferty H, Rail S~X, Raja S, Rajan C, Rajbhandari B, Rakhmanov M, Ramirez
  K~E, Ramirez T~D, Ramos-Buades A, Rana J, Rao K, Rapagnani P, Rapol U~D,
  Ratto B, Raymond V, Razzano M, Read J, Regimbau T, Rei L, Reid S, Reitze D~H,
  Rettegno P, Ricci F, Richardson C~J, Richardson J~W, Richardson L, Ricker
  P~M, Riemenschneider G, Riles K, Rizzo M, Robertson N~A, Robinet F, Rocchi A,
  Rocha J~A, Rodriguez S, Rodriguez-Soto R~D, Rolland L, Rollins J~G, Roma V~J,
  Romanelli M, Romano R, Romel C~L, Romero A, Romero-Shaw I~M, Romie J~H,
  Ronchini S, Rose C~A, Rose D, Rose K, Rosell M~J~B,
  Rosi\ifmmode~\acute{n}\else \'{n}\fi{}ska D, Rosofsky S~G, Ross M~P, Rowan S,
  Rowlinson S~J, Roy S, Roy S, Ruggi P, Ryan K, Sachdev S, Sadecki T, Sadiq J,
  Sakellariadou M, Salafia O~S, Salconi L, Saleem M, Samajdar A, Sanchez E~J,
  Sanchez J~H, Sanchez L~E, Sanchis-Gual N, Sanders J~R, Sandles L, Santiago
  K~A, Santos E, Saravanan T~R, Sarin N, Sassolas B, Sathyaprakash B~S, Sauter
  O, Savage R~L, Savant V, Sawant D, Sayah S, Schaetzl D, Schale P, Scheel M,
  Scheuer J, Schindler-Tyka A, Schmidt P, Schnabel R, Schofield R~M~S,
  Sch\"onbeck A, Schreiber E, Schulte B~W, Schutz B~F, Schwarm O, Schwartz E,
  Scott J, Scott S~M, Seglar-Arroyo M, Seidel E, Sellers D, Sengupta A~S,
  Sennett N, Sentenac D, Sequino V, Sergeev A, Setyawati Y, Shaffer T, Shahriar
  M~S, Sharifi S, Sharma A, Sharma P, Shawhan P, Shen H, Shikauchi M, Shink R,
  Shoemaker D~H, Shoemaker D~M, Shukla K, ShyamSundar S, Sieniawska M, Sigg D,
  Singer L~P, Singh D, Singh N, Singha A, Singhal A, Sintes A~M, Sipala V,
  Skliris V, Slagmolen B~J~J, Slaven-Blair T~J, Smetana J, Smith J~R, Smith
  R~J~E, Somala S~N, Son E~J, Soni K, Soni S, Sorazu B, Sordini V, Sorrentino
  F, Sorrentino N, Soulard R, Souradeep T, Sowell E, Spencer A~P, Spera M,
  Srivastava A~K, Srivastava V, Staats K, Stachie C, Steer D~A, Steinhoff J,
  Steinke M, Steinlechner J, Steinlechner S, Steinmeyer D, Stevenson S~P,
  Stolle-McAllister G, Stops D~J, Stover M, Strain K~A, Stratta G, Strunk A,
  Sturani R, Stuver A~L, S\"udbeck J, Sudhagar S, Sudhir V, Suh H~G,
  Summerscales T~Z, Sun H, Sun L, Sunil S, Sur A, Suresh J, Sutton P~J,
  Swinkels B~L, Szczepa\ifmmode~\acute{n}\else \'{n}\fi{}czyk M~J, Tacca M,
  Tait S~C, Talbot C, Tanasijczuk A~J, Tanner D~B, Tao D, Tapia A, Tapia
  San~Martin E~N, Tasson J~D, Taylor R, Tenorio R, Terkowski L,
  Thirugnanasambandam M~P, Thomas L~M, Thomas M, Thomas P, Thompson J~E,
  Thondapu S~R, Thorne K~A, Thrane E, Tiwari S, Tiwari S, Tiwari V, Toland K,
  Tolley A~E, Tonelli M, Tornasi Z, Torres-Forn\'e A, Torrie C~I, e~Melo I~T,
  T\"oyr\"a D, Tran A~T, Trapananti A, Travasso F, Traylor G, Tringali M~C,
  Tripathee A, Trovato A, Trudeau R~J, Tsai D~S, Tsang K~W, Tse M, Tso R,
  Tsukada L, Tsuna D, Tsutsui T, Turconi M, Ubhi A~S, Udall R~P, Ueno K,
  Ugolini D, Unnikrishnan C~S, Urban A~L, Usman S~A, Utina A~C, Vahlbruch H,
  Vajente G, Vajpeyi A, Valdes G, Valentini M, Valsan V, van Bakel N, van
  Beuzekom M, van~den Brand J~F~J, Van Den~Broeck C, Vander-Hyde D~C, van~der
  Schaaf L, van Heijningen J~V, Vardaro M, Vargas A~F, Varma V, Vass S,
  Vas\'uth M, Vecchio A, Vedovato G, Veitch J, Veitch P~J, Venkateswara K,
  Venneberg J, Venugopalan G, Verkindt D, Verma Y, Veske D, Vetrano F, Vicer\'e
  A, Viets A~D, Vijaykumar A, Villa-Ortega V, Vinet J~Y, Vitale S, Vo T, Vocca
  H, Vorvick C, Vyatchanin S~P, Wade A~R, Wade L~E, Wade M, Walet R~C, Walker
  M, Wallace G~S, Wallace L, Walsh S, Wang J~Z, Wang S, Wang W~H, Wang Y~F,
  Ward R~L, Warner J, Was M, Washington N~Y, Watchi J, Weaver B, Wei L, Weinert
  M, Weinstein A~J, Weiss R, Wellmann F, Wen L, We\ss{}els P, Westhouse J~W,
  Wette K, Whelan J~T, White D~D, White L~V, Whiting B~F, Whittle C, Wilken
  D~M, Williams D, Williams M~J, Williamson A~R, Willis J~L, Willke B, Wilson
  D~J, Wimmer M~H, Winkler W, Wipf C~C, Woan G, Woehler J, Wofford J~K, Wong
  I~C~F, Wrangel J, Wright J~L, Wu D~S, Wysocki D~M, Xiao L, Yamamoto H, Yang
  L, Yang Y, Yang Z, Yap M~J, Yeeles D~W, Yoon A, Yu H, Yu H, Yuen S~H~R,
  Zadro\ifmmode~\dot{z}\else \.{z}\fi{}ny A, Zanolin M, Zelenova T, Zendri J~P,
  Zevin M, Zhang J, Zhang L, Zhang R, Zhang T, Zhao C, Zhao G, Zheng Y, Zhou M,
  Zhou Z, Zhu X~J, Zimmerman A~B, Zlochower Y, Zucker M~E and Zweizig J (LIGO
  Scientific Collaboration and Virgo Collaboration) 2021 {\em Phys. Rev. X\/}
  {\bf 11}(2) 021053
  \urlprefix\url{https://link.aps.org/doi/10.1103/PhysRevX.11.021053}

\bibitem{AbEA2021a}
Collaboration T~L~S, the Virgo~Collaboration, the KAGRA~Collaboration, Abbott
  R, Abbott T~D, Acernese F, Ackley K, Adams C, Adhikari N, Adhikari R~X, Adya
  V~B, Affeldt C, Agarwal D, Agathos M, Agatsuma K, Aggarwal N, Aguiar O~D,
  Aiello L, Ain A, Ajith P, Akcay S, Akutsu T, Albanesi S, Allocca A, Altin
  P~A, Amato A, Anand C, Anand S, Ananyeva A, Anderson S~B, Anderson W~G, Ando
  M, Andrade T, Andres N, Andrić T, Angelova S~V, Ansoldi S, Antelis J~M,
  Antier S, Appert S, Arai K, Arai K, Arai Y, Araki S, Araya A, Araya M~C,
  Areeda J~S, Arène M, Aritomi N, Arnaud N, Arogeti M, Aronson S~M, Arun K~G,
  Asada H, Asali Y, Ashton G, Aso Y, Assiduo M, Aston S~M, Astone P, Aubin F,
  Austin C, Babak S, Badaracco F, Bader M~K~M, Badger C, Bae S, Bae Y, Baer
  A~M, Bagnasco S, Bai Y, Baiotti L, Baird J, Bajpai R, Ball M, Ballardin G,
  Ballmer S~W, Balsamo A, Baltus G, Banagiri S, Bankar D, Barayoga J~C,
  Barbieri C, Barish B~C, Barker D, Barneo P, Barone F, Barr B, Barsotti L,
  Barsuglia M, Barta D, Bartlett J, Barton M~A, Bartos I, Bassiri R, Basti A,
  Bawaj M, Bayley J~C, Baylor A~C, Bazzan M, Bécsy B, Bedakihale V~M, Bejger
  M, Belahcene I, Benedetto V, Beniwal D, Bennett T~F, Bentley J~D, BenYaala M,
  Bergamin F, Berger B~K, Bernuzzi S, Berry C~P~L, Bersanetti D, Bertolini A,
  Betzwieser J, Beveridge D, Bhandare R, Bhardwaj U, Bhattacharjee D, Bhaumik
  S, Bilenko I~A, Billingsley G, Bini S, Birney R, Birnholtz O, Biscans S,
  Bischi M, Biscoveanu S, Bisht A, Biswas B, Bitossi M, Bizouard M~A, Blackburn
  J~K, Blair C~D, Blair D~G, Blair R~M, Bobba F, Bode N, Boer M, Bogaert G,
  Boldrini M, Bonavena L~D, Bondu F, Bonilla E, Bonnand R, Booker P, Boom B~A,
  Bork R, Boschi V, Bose N, Bose S, Bossilkov V, Boudart V, Bouffanais Y, Bozzi
  A, Bradaschia C, Brady P~R, Bramley A, Branch A, Branchesi M, Brandt J, Brau
  J~E, Breschi M, Briant T, Briggs J~H, Brillet A, Brinkmann M, Brockill P,
  Brooks A~F, Brooks J, Brown D~D, Brunett S, Bruno G, Bruntz R, Bryant J,
  Bulik T, Bulten H~J, Buonanno A, Buscicchio R, Buskulic D, Buy C, Byer R~L,
  Davies G~S~C, Cadonati L, Cagnoli G, Cahillane C, Bustillo J~C, Callaghan
  J~D, Callister T~A, Calloni E, Cameron J, Camp J~B, Canepa M, Canevarolo S,
  Cannavacciuolo M, Cannon K~C, Cao H, Cao Z, Capocasa E, Capote E, Carapella
  G, Carbognani F, Carlin J~B, Carney M~F, Carpinelli M, Carrillo G, Carullo G,
  Carver T~L, Diaz J~C, Casentini C, Castaldi G, Caudill S, Cavaglià M,
  Cavalier F, Cavalieri R, Ceasar M, Cella G, Cerdá-Durán P, Cesarini E,
  Chaibi W, Chakravarti K, Subrahmanya S~C, Champion E, Chan C~H, Chan C, Chan
  C~L, Chan K, Chan M, Chandra K, Chanial P, Chao S, Chapman-Bird C~E~A,
  Charlton P, Chase E~A, Chassande-Mottin E, Chatterjee C, Chatterjee D,
  Chatterjee D, Chaturvedi M, Chaty S, Chatziioannou K, Chen C, Chen H~Y, Chen
  J, Chen K, Chen X, Chen Y~B, Chen Y~R, Chen Z, Cheng H, Cheong C~K, Cheung
  H~Y, Chia H~Y, Chiadini F, Chiang C~Y, Chiarini G, Chierici R, Chincarini A,
  Chiofalo M~L, Chiummo A, Cho G, Cho H~S, Choudhary R~K, Choudhary S,
  Christensen N, Chu H, Chu Q, Chu Y~K, Chua S, Chung K~W, Ciani G, Ciecielag
  P, Cieślar M, Cifaldi M, Ciobanu A~A, Ciolfi R, Cipriano F, Cirone A, Clara
  F, Clark E~N, Clark J~A, Clarke L, Clearwater P, Clesse S, Cleva F, Coccia E,
  Codazzo E, Cohadon P~F, Cohen D~E, Cohen L, Colleoni M, Collette C~G, Colombo
  A, Colpi M, Compton C~M, au2 M~C~J, Conti L, Cooper S~J, Corban P, Corbitt
  T~R, Cordero-Carrión I, Corezzi S, Corley K~R, Cornish N, Corre D, Corsi A,
  Cortese S, Costa C~A, Cotesta R, Coughlin M~W, Coulon J~P, Countryman S~T,
  Cousins B, Couvares P, Coward D~M, Cowart M~J, Coyne D~C, Coyne R, Creighton
  J~D~E, Creighton T~D, Criswell A~W, Croquette M, Crowder S~G, Cudell J~R,
  Cullen T~J, Cumming A, Cummings R, Cunningham L, Cuoco E, Curyło M, Dabadie
  P, Canton T~D, Dall'Osso S, Dálya G, Dana A, DaneshgaranBajastani L~M,
  D'Angelo B, Danila B, Danilishin S, D'Antonio S, Danzmann K, Darsow-Fromm C,
  Dasgupta A, Datrier L~E~H, Datta S, Dattilo V, Dave I, Davier M, Davis D,
  Davis M~C, Daw E~J, de~Alarcón P~F, Dean R, DeBra D, Deenadayalan M,
  Degallaix J, Laurentis M~D, Deléglise S, Favero V~D, Lillo F~D, Lillo N~D,
  Pozzo W~D, DeMarchi L~M, Matteis F~D, D'Emilio V, Demos N, Dent T, Depasse A,
  Pietri R~D, Rosa R~D, Rossi C~D, DeSalvo R, Simone R~D, Dhurandhar S, Díaz
  M~C, au2 M~D~O~J, Didio N~A, Dietrich T, Fiore L~D, Fronzo C~D, Giorgio C~D,
  Giovanni F~D, Giovanni M~D, Girolamo T~D, Lieto A~D, Ding B, Pace S~D, Palma
  I~D, Renzo F~D, Divakarla A~K, Dmitriev A, Doctor Z, D'Onofrio L, Donovan F,
  Dooley K~L, Doravari S, Dorrington I, Drago M, Driggers J~C, Drori Y, Ducoin
  J~G, Dupej P, Durante O, D'Urso D, Duverne P~A, Dwyer S~E, Eassa C, Easter
  P~J, Ebersold M, Eckhardt T, Eddolls G, Edelman B, Edo T~B, Edy O, Effler A,
  Eguchi S, Eichholz J, Eikenberry S~S, Eisenmann M, Eisenstein R~A, Ejlli A,
  Engelby E, Enomoto Y, Errico L, Essick R~C, Estellés H, Estevez D, Etienne
  Z, Etzel T, Evans M, Evans T~M, Ewing B~E, Fafone V, Fair H, Fairhurst S,
  Farah A~M, Farinon S, Farr B, Farr W~M, Farrow N~W, Fauchon-Jones E~J, Favaro
  G, Favata M, Fays M, Fazio M, Feicht J, Fejer M~M, Fenyvesi E, Ferguson D~L,
  Fernandez-Galiana A, Ferrante I, Ferreira T~A, Fidecaro F, Figura P, Fiori I,
  Fishbach M, Fisher R~P, Fittipaldi R, Fiumara V, Flaminio R, Floden E, Fong
  H, Font J~A, Fornal B, Forsyth P~W~F, Franke A, Frasca S, Frasconi F,
  Frederick C, Freed J~P, Frei Z, Freise A, Frey R, Fritschel P, Frolov V~V,
  Fronzé G~G, Fujii Y, Fujikawa Y, Fukunaga M, Fukushima M, Fulda P, Fyffe M,
  Gabbard H~A, Gabella W~E, Gadre B~U, Gair J~R, Gais J, Galaudage S, Gamba R,
  Ganapathy D, Ganguly A, Gao D, Gaonkar S~G, Garaventa B, García F,
  García-Núñez C, García-Quirós C, Garufi F, Gateley B, Gaudio S, Gayathri
  V, Ge G~G, Gemme G, Gennai A, George J, George R~N, Gerberding O, Gergely L,
  Gewecke P, Ghonge S, Ghosh A, Ghosh A, Ghosh S, Ghosh S, Giacomazzo B,
  Giacoppo L, Giaime J~A, Giardina K~D, Gibson D~R, Gier C, Giesler M, Giri P,
  Gissi F, Glanzer J, Gleckl A~E, Godwin P, Goetz E, Goetz R, Gohlke N, Golomb
  J, Goncharov B, González G, Gopakumar A, Gosselin M, Gouaty R, Gould D~W,
  Grace B, Grado A, Granata M, Granata V, Grant A, Gras S, Grassia P, Gray C,
  Gray R, Greco G, Green A~C, Green R, Gretarsson A~M, Gretarsson E~M, Griffith
  D, Griffiths W, Griggs H~L, Grignani G, Grimaldi A, Grimm S~J, Grote H,
  Grunewald S, Gruning P, Guerra D, Guidi G~M, Guimaraes A~R, Guixé G, Gulati
  H~K, Guo H~K, Guo Y, Gupta A, Gupta A, Gupta P, Gustafson E~K, Gustafson R,
  Guzman F, Ha S, Haegel L, Hagiwara A, Haino S, Halim O, Hall E~D, Hamilton
  E~Z, Hammond G, Han W~B, Haney M, Hanks J, Hanna C, Hannam M~D, Hannuksela O,
  Hansen H, Hansen T~J, Hanson J, Harder T, Hardwick T, Haris K, Harms J, Harry
  G~M, Harry I~W, Hartwig D, Hasegawa K, Haskell B, Hasskew R~K, Haster C~J,
  Hattori K, Haughian K, Hayakawa H, Hayama K, Hayes F~J, Healy J, Heidmann A,
  Heidt A, Heintze M~C, Heinze J, Heinzel J, Heitmann H, Hellman F, Hello P,
  Helmling-Cornell A~F, Hemming G, Hendry M, Heng I~S, Hennes E, Hennig J,
  Hennig M~H, Hernandez A~G, Vivanco F~H, Heurs M, Hild S, Hill P, Himemoto Y,
  Hines A~S, Hiranuma Y, Hirata N, Hirose E, Hochheim S, Hofman D, Hohmann J~N,
  Holcomb D~G, Holland N~A, Holley-Bockelmann K, Hollows I~J, Holmes Z~J, Holt
  K, Holz D~E, Hong Z, Hopkins P, Hough J, Hourihane S, Howell E~J, Hoy C~G,
  Hoyland D, Hreibi A, Hsieh B~H, Hsu Y, Huang G~Z, Huang H~Y, Huang P, Huang
  Y~C, Huang Y~J, Huang Y, Hübner M~T, Huddart A~D, Hughey B, Hui D~C~Y, Hui
  V, Husa S, Huttner S~H, Huxford R, Huynh-Dinh T, Ide S, Idzkowski B, Iess A,
  Ikenoue B, Imam S, Inayoshi K, Ingram C, Inoue Y, Ioka K, Isi M, Isleif K,
  Ito K, Itoh Y, Iyer B~R, Izumi K, JaberianHamedan V, Jacqmin T, Jadhav S~J,
  Jadhav S~P, James A~L, Jan A~Z, Jani K, Janquart J, Janssens K, Janthalur
  N~N, Jaranowski P, Jariwala D, Jaume R, Jenkins A~C, Jenner K, Jeon C, Jeunon
  M, Jia W, Jin H~B, Johns G~R, Johnson-McDaniel N~K, Jones A~W, Jones D~I,
  Jones J~D, Jones P, Jones R, Jonker R~J~G, Ju L, Jung P, Jung K, Junker J,
  Juste V, Kaihotsu K, Kajita T, Kakizaki M, Kalaghatgi C~V, Kalogera V, Kamai
  B, Kamiizumi M, Kanda N, Kandhasamy S, Kang G, Kanner J~B, Kao Y, Kapadia
  S~J, Kapasi D~P, Karat S, Karathanasis C, Karki S, Kashyap R, Kasprzack M,
  Kastaun W, Katsanevas S, Katsavounidis E, Katzman W, Kaur T, Kawabe K,
  Kawaguchi K, Kawai N, Kawasaki T, Kéfélian F, Keitel D, Key J~S, Khadka S,
  Khalili F~Y, Khan S, Khazanov E~A, Khetan N, Khursheed M, Kijbunchoo N, Kim
  C, Kim J~C, Kim J, Kim K, Kim W~S, Kim Y~M, Kimball C, Kimura N,
  Kinley-Hanlon M, Kirchhoff R, Kissel J~S, Kita N, Kitazawa H, Kleybolte L,
  Klimenko S, Knee A~M, Knowles T~D, Knyazev E, Koch P, Koekoek G, Kojima Y,
  Kokeyama K, Koley S, Kolitsidou P, Kolstein M, Komori K, Kondrashov V, Kong
  A~K~H, Kontos A, Koper N, Korobko M, Kotake K, Kovalam M, Kozak D~B, Kozakai
  C, Kozu R, Kringel V, Krishnendu N~V, Królak A, Kuehn G, Kuei F, Kuijer P,
  Kulkarni S, Kumar A, Kumar P, Kumar R, Kumar R, Kume J, Kuns K, Kuo C, Kuo
  H~S, Kuromiya Y, Kuroyanagi S, Kusayanagi K, Kuwahara S, Kwak K, Lagabbe P,
  Laghi D, Lalande E, Lam T~L, Lamberts A, Landry M, Lane B~B, Lang R~N, Lange
  J, Lantz B, Rosa I~L, Lartaux-Vollard A, Lasky P~D, Laxen M, Lazzarini A,
  Lazzaro C, Leaci P, Leavey S, Lecoeuche Y~K, Lee H~K, Lee H~M, Lee H~W, Lee
  J, Lee K, Lee R, Lehmann J, Lemaître A, Leonardi M, Leroy N, Letendre N,
  Levesque C, Levin Y, Leviton J~N, Leyde K, Li A~K~Y, Li B, Li J, Li K~L, Li
  T~G~F, Li X, Lin C~Y, Lin F~K, Lin F~L, Lin H~L, Lin L~C~C, Linde F, Linker
  S~D, Linley J~N, Littenberg T~B, Liu G~C, Liu J, Liu K, Liu X, Llamas F,
  Llorens-Monteagudo M, Lo R~K~L, Lockwood A, Loh M, London L~T, Longo A, Lopez
  D, Portilla M~L, Lorenzini M, Loriette V, Lormand M, Losurdo G, Lott T~P,
  Lough J~D, Lousto C~O, Lovelace G, Lucaccioni J~F, Lück H, Lumaca D,
  Lundgren A~P, Luo L~W, Lynam J~E, Macas R, MacInnis M, Macleod D~M, MacMillan
  I~A~O, Macquet A, Hernandez I~M, Magazzù C, Magee R~M, Maggiore R, Magnozzi
  M, Mahesh S, Majorana E, Makarem C, Maksimovic I, Maliakal S, Malik A, Man N,
  Mandic V, Mangano V, Mango J~L, Mansell G~L, Manske M, Mantovani M, Mapelli
  M, Marchesoni F, Marchio M, Marion F, Mark Z, Márka S, Márka Z, Markakis C,
  Markosyan A~S, Markowitz A, Maros E, Marquina A, Marsat S, Martelli F, Martin
  I~W, Martin R~M, Martinez M, Martinez V~A, Martinez V, Martinovic K, Martynov
  D~V, Marx E~J, Masalehdan H, Mason K, Massera E, Masserot A, Massinger T~J,
  Masso-Reid M, Mastrogiovanni S, Matas A, Mateu-Lucena M, Matichard F,
  Matiushechkina M, Mavalvala N, McCann J~J, McCarthy R, McClelland D~E,
  McClincy P~K, McCormick S, McCuller L, McGhee G~I, McGuire S~C, McIsaac C,
  McIver J, McRae T, McWilliams S~T, Meacher D, Mehmet M, Mehta A~K, Meijer Q,
  Melatos A, Melchor D~A, Mendell G, Menendez-Vazquez A, Menoni C~S, Mercer
  R~A, Mereni L, Merfeld K, Merilh E~L, Merritt J~D, Merzougui M, Meshkov S,
  Messenger C, Messick C, Meyers P~M, Meylahn F, Mhaske A, Miani A, Miao H,
  Michaloliakos I, Michel C, Michimura Y, Middleton H, Milano L, Miller A~L,
  Miller A, Miller B, Millhouse M, Mills J~C, Milotti E, Minazzoli O, Minenkov
  Y, Mio N, Mir L~M, Miravet-Tenés M, Mishra C, Mishra T, Mistry T, Mitra S,
  Mitrofanov V~P, Mitselmakher G, Mittleman R, Miyakawa O, Miyamoto A, Miyazaki
  Y, Miyo K, Miyoki S, Mo G, Modafferi L~M, Moguel E, Mogushi K, Mohapatra
  S~R~P, Mohite S~R, Molina I, Molina-Ruiz M, Mondin M, Montani M, Moore C~J,
  Moraru D, Morawski F, More A, Moreno C, Moreno G, Mori Y, Morisaki S,
  Moriwaki Y, Morrás G, Mours B, Mow-Lowry C~M, Mozzon S, Muciaccia F,
  Mukherjee A, Mukherjee D, Mukherjee S, Mukherjee S, Mukherjee S, Mukund N,
  Mullavey A, Munch J, Muñiz E~A, Murray P~G, Musenich R, Muusse S, Nadji S~L,
  Nagano K, Nagano S, Nagar A, Nakamura K, Nakano H, Nakano M, Nakashima R,
  Nakayama Y, Napolano V, Nardecchia I, Narikawa T, Naticchioni L, Nayak B,
  Nayak R~K, Negishi R, Neil B~F, Neilson J, Nelemans G, Nelson T~J~N, Nery M,
  Neubauer P, Neunzert A, Ng K~Y, Ng S~W~S, Nguyen C, Nguyen P, Nguyen T, Quynh
  L~N, Ni W~T, Nichols S~A, Nishizawa A, Nissanke S, Nitoglia E, Nocera F,
  Norman M, North C, Nozaki S, Siles J~F~N, Nuttall L~K, Oberling J, O'Brien
  B~D, Obuchi Y, O'Dell J, Oelker E, Ogaki W, Oganesyan G, Oh J~J, Oh K, Oh
  S~H, Ohashi M, Ohishi N, Ohkawa M, Ohme F, Ohta H, Okada M~A, Okutani Y,
  Okutomi K, Olivetto C, Oohara K, Ooi C, Oram R, O'Reilly B, Ormiston R~G,
  Ormsby N~D, Ortega L~F, O'Shaughnessy R, O'Shea E, Oshino S, Ossokine S,
  Osthelder C, Otabe S, Ottaway D~J, Overmier H, Pace A~E, Pagano G, Page M~A,
  Pagliaroli G, Pai A, Pai S~A, Palamos J~R, Palashov O, Palomba C, Pan H, Pan
  K, Panda P~K, Pang H, Pang P~T~H, Pankow C, Pannarale F, Pant B~C, Panther
  F~H, Paoletti F, Paoli A, Paolone A, Parisi A, Park H, Park J, Parker W,
  Pascucci D, Pasqualetti A, Passaquieti R, Passuello D, Patel M, Pathak M,
  Patricelli B, Patron A~S, Paul S, Payne E, Pedraza M, Pegoraro M, Pele A,
  Arellano F~E~P, Penn S, Perego A, Pereira A, Pereira T, Perez C~J, Périgois
  C, Perkins C~C, Perreca A, Perriès S, Petermann J, Petterson D, Pfeiffer
  H~P, Pham K~A, Phukon K~S, Piccinni O~J, Pichot M, Piendibene M, Piergiovanni
  F, Pierini L, Pierro V, Pillant G, Pillas M, Pilo F, Pinard L, Pinto I~M,
  Pinto M, Piotrzkowski B, Piotrzkowski K, Pirello M, Pitkin M~D, Placidi E,
  Planas L, Plastino W, Pluchar C, Poggiani R, Polini E, Pong D~Y~T, Ponrathnam
  S, Popolizio P, Porter E~K, Poulton R, Powell J, Pracchia M, Pradier T,
  Prajapati A~K, Prasai K, Prasanna R, Pratten G, Principe M, Prodi G~A,
  Prokhorov L, Prosposito P, Prudenzi L, Puecher A, Punturo M, Puosi F, Puppo
  P, Pürrer M, Qi H, Quetschke V, Quitzow-James R, Qutob N, Raab F~J,
  Raaijmakers G, Radkins H, Radulesco N, Raffai P, Rail S~X, Raja S, Rajan C,
  Ramirez K~E, Ramirez T~D, Ramos-Buades A, Rana J, Rapagnani P, Rapol U~D, Ray
  A, Raymond V, Raza N, Razzano M, Read J, Rees L~A, Regimbau T, Rei L, Reid S,
  Reid S~W, Reitze D~H, Relton P, Renzini A, Rettegno P, Reza A, Rezac M, Ricci
  F, Richards D, Richardson J~W, Richardson L, Riemenschneider G, Riles K,
  Rinaldi S, Rink K, Rizzo M, Robertson N~A, Robie R, Robinet F, Rocchi A,
  Rodriguez S, Rolland L, Rollins J~G, Romanelli M, Romano R, Romel C~L,
  Romero-Rodríguez A, Romero-Shaw I~M, Romie J~H, Ronchini S, Rosa L, Rose
  C~A, Rosińska D, Ross M~P, Rowan S, Rowlinson S~J, Roy S, Roy S, Roy S,
  Rozza D, Ruggi P, Ruiz-Rocha K, Ryan K, Sachdev S, Sadecki T, Sadiq J, Sago
  N, Saito S, Saito Y, Sakai K, Sakai Y, Sakellariadou M, Sakuno Y, Salafia
  O~S, Salconi L, Saleem M, Salemi F, Samajdar A, Sanchez E~J, Sanchez J~H,
  Sanchez L~E, Sanchis-Gual N, Sanders J~R, Sanuy A, Saravanan T~R, Sarin N,
  Sassolas B, Satari H, Sathyaprakash B~S, Sato S, Sato T, Sauter O, Savage
  R~L, Sawada T, Sawant D, Sawant H~L, Sayah S, Schaetzl D, Scheel M, Scheuer
  J, Schiworski M, Schmidt P, Schmidt S, Schnabel R, Schneewind M, Schofield
  R~M~S, Schönbeck A, Schulte B~W, Schutz B~F, Schwartz E, Scott J, Scott S~M,
  Seglar-Arroyo M, Sekiguchi T, Sekiguchi Y, Sellers D, Sengupta A~S, Sentenac
  D, Seo E~G, Sequino V, Sergeev A, Setyawati Y, Shaffer T, Shahriar M~S, Shams
  B, Shao L, Sharma A, Sharma P, Shawhan P, Shcheblanov N~S, Shibagaki S,
  Shikauchi M, Shimizu R, Shimoda T, Shimode K, Shinkai H, Shishido T, Shoda A,
  Shoemaker D~H, Shoemaker D~M, ShyamSundar S, Sieniawska M, Sigg D, Singer
  L~P, Singh D, Singh N, Singha A, Sintes A~M, Sipala V, Skliris V, Slagmolen
  B~J~J, Slaven-Blair T~J, Smetana J, Smith J~R, Smith R~J~E, Soldateschi J,
  Somala S~N, Somiya K, Son E~J, Soni K, Soni S, Sordini V, Sorrentino F,
  Sorrentino N, Sotani H, Soulard R, Souradeep T, Sowell E, Spagnuolo V,
  Spencer A~P, Spera M, Srinivasan R, Srivastava A~K, Srivastava V, Staats K,
  Stachie C, Steer D~A, Steinhoff J, Steinlechner J, Steinlechner S, Stevenson
  S~P, Stops D~J, Stover M, Strain K~A, Strang L~C, Stratta G, Strunk A,
  Sturani R, Stuver A~L, Sudhagar S, Sudhir V, Sugimoto R, Suh H~G, Sullivan
  A~G, Sullivan J~M, Summerscales T~Z, Sun H, Sun L, Sunil S, Sur A, Suresh J,
  Sutton P~J, Suzuki T, Suzuki T, Swinkels B~L, Szczepańczyk M~J, Szewczyk P,
  Tacca M, Tagoshi H, Tait S~C, Takahashi H, Takahashi R, Takamori A, Takano S,
  Takeda H, Takeda M, Talbot C~J, Talbot C, Tanaka H, Tanaka K, Tanaka K,
  Tanaka T, Tanaka T, Tanasijczuk A~J, Tanioka S, Tanner D~B, Tao D, Tao L,
  Martín E~N~T~S, Taranto C, Tasson J~D, Telada S, Tenorio R, Terhune J~E,
  Terkowski L, Thirugnanasambandam M~P, Thomas L, Thomas M, Thomas P, Thompson
  J~E, Thondapu S~R, Thorne K~A, Thrane E, Tiwari S, Tiwari S, Tiwari V,
  Toivonen A~M, Toland K, Tolley A~E, Tomaru T, Tomigami Y, Tomura T, Tonelli
  M, Torres-Forné A, Torrie C~I, e~Melo I~T, Töyrä D, Trapananti A, Travasso
  F, Traylor G, Trevor M, Tringali M~C, Tripathee A, Troiano L, Trovato A,
  Trozzo L, Trudeau R~J, Tsai D~S, Tsai D, Tsang K~W, Tsang T, Tsao J~S, Tse M,
  Tso R, Tsubono K, Tsuchida S, Tsukada L, Tsuna D, Tsutsui T, Tsuzuki T,
  Turbang K, Turconi M, Tuyenbayev D, Ubhi A~S, Uchikata N, Uchiyama T, Udall
  R~P, Ueda A, Uehara T, Ueno K, Ueshima G, Unnikrishnan C~S, Uraguchi F, Urban
  A~L, Ushiba T, Utina A, Vahlbruch H, Vajente G, Vajpeyi A, Valdes G,
  Valentini M, Valsan V, van Bakel N, van Beuzekom M, van~den Brand J~F~J,
  Broeck C~V~D, Vander-Hyde D~C, van~der Schaaf L, van Heijningen J~V, Vanosky
  J, van Putten M~H~P~M, van Remortel N, Vardaro M, Vargas A~F, Varma V,
  Vasúth M, Vecchio A, Vedovato G, Veitch J, Veitch P~J, Venneberg J,
  Venugopalan G, Verkindt D, Verma P, Verma Y, Veske D, Vetrano F, Viceré A,
  Vidyant S, Viets A~D, Vijaykumar A, Villa-Ortega V, Vinet J~Y, Virtuoso A,
  Vitale S, Vo T, Vocca H, von Reis E~R~G, von Wrangel J~S~A, Vorvick C,
  Vyatchanin S~P, Wade L~E, Wade M, Wagner K~J, Walet R~C, Walker M, Wallace
  G~S, Wallace L, Walsh S, Wang J, Wang J~Z, Wang W~H, Ward R~L, Warner J, Was
  M, Washimi T, Washington N~Y, Watchi J, Weaver B, Webster S~A, Weinert M,
  Weinstein A~J, Weiss R, Weller C~M, Weller R~A, Wellmann F, Wen L, Weßels P,
  Wette K, Whelan J~T, White D~D, Whiting B~F, Whittle C, Wilken D, Williams D,
  Williams M~J, Williams N, Williamson A~R, Willis J~L, Willke B, Wilson D~J,
  Winkler W, Wipf C~C, Wlodarczyk T, Woan G, Woehler J, Wofford J~K, Wong
  I~C~F, Wu C, Wu D~S, Wu H, Wu S, Wysocki D~M, Xiao L, Xu W~R, Yamada T,
  Yamamoto H, Yamamoto K, Yamamoto K, Yamamoto T, Yamashita K, Yamazaki R, Yang
  F~W, Yang L, Yang Y, Yang Y, Yang Z, Yap M~J, Yeeles D~W, Yelikar A~B, Ying
  M, Yokogawa K, Yokoyama J, Yokozawa T, Yoo J, Yoshioka T, Yu H, Yu H,
  Yuzurihara H, Zadrożny A, Zanolin M, Zeidler S, Zelenova T, Zendri J~P,
  Zevin M, Zhan M, Zhang H, Zhang J, Zhang L, Zhang T, Zhang Y, Zhao C, Zhao G,
  Zhao Y, Zhao Y, Zheng Y, Zhou R, Zhou Z, Zhu X~J, Zhu Z~H, Zimmerman A~B,
  Zlochower Y, Zucker M~E and Zweizig J 2021 {GWTC-3: Compact Binary
  Coalescences Observed by LIGO and Virgo During the Second Part of the Third
  Observing Run} \urlprefix\url{https://arxiv.org/abs/2111.03606}

\bibitem{AkEA2018}
Akutsu T, Ando M, Arai K, Arai Y, Araki S, Araya A, Aritomi N, Asada H, Aso Y,
  Atsuta S, Awai K, Bae S, Baiotti L, Barton M~A, Cannon K, Capocasa E, Chen
  C~S, Chiu T~W, Cho K, Chu Y~K, Craig K, Creus W, Doi K, Eda K, Enomoto Y,
  Flaminio R, Fujii Y, Fujimoto M~K, Fukunaga M, Fukushima M, Furuhata T, Haino
  S, Hasegawa K, Hashino K, Hayama K, Hirobayashi S, Hirose E, Hsieh B~H, Huang
  C~Z, Ikenoue B, Inoue Y, Ioka K, Itoh Y, Izumi K, Kaji T, Kajita T, Kakizaki
  M, Kamiizumi M, Kanbara S, Kanda N, Kanemura S, Kaneyama M, Kang G, Kasuya J,
  Kataoka Y, Kawai N, Kawamura S, Kawasaki T, Kim C, Kim J, Kim J~C, Kim W~S,
  Kim Y~M, Kimura N, Kinugawa T, Kirii S, Kitaoka Y, Kitazawa H, Kojima Y,
  Kokeyama K, Komori K, Kong A~K~H, Kotake K, Kozu R, Kumar R, Kuo H~S,
  Kuroyanagi S, Lee H~K, Lee H~M, Lee H~W, Leonardi M, Lin C~Y, Lin F~L, Liu
  G~C, Liu Y, Majorana E, Mano S, Marchio M, Matsui T, Matsushima F, Michimura
  Y, Mio N, Miyakawa O, Miyamoto A, Miyamoto T, Miyo K, Miyoki S, Morii W,
  Morisaki S, Moriwaki Y, Morozumi T, Musha M, Nagano K, Nagano S, Nakamura K,
  Nakamura T, Nakano H, Nakano M, Nakao K, Narikawa T, Naticchioni L, Quynh
  L~N, Ni W~T, Nishizawa A, Obuchi Y, Ochi T, Oh J~J, Oh S~H, Ohashi M, Ohishi
  N, Ohkawa M, Okutomi K, Ono K, Oohara K, Ooi C~P, Pan S~S, Park J, Arellano
  F~E~P, Pinto I, Sago N, Saijo M, Saitou S, Saito Y, Sakai K, Sakai Y, Sakai
  Y, Sasai M, Sasaki M, Sasaki Y, Sato S, Sato N, Sato T, Sekiguchi Y, Seto N,
  Shibata M, Shimoda T, Shinkai H, Shishido T, Shoda A, Somiya K, Son E~J,
  Suemasa A, Suzuki T, Suzuki T, Tagoshi H, Tahara H, Takahashi H, Takahashi R,
  Takamori A, Takeda H, Tanaka H, Tanaka K, Tanaka T, Tanioka S, Martin
  E~N~T~S, Tatsumi D, Tomaru T, Tomura T, Travasso F, Tsubono K, Tsuchida S,
  Uchikata N, Uchiyama T, Uehara T, Ueki S, Ueno K, Uraguchi F, Ushiba T, van
  Putten M~H~P~M, Vocca H, Wada S, Wakamatsu T, Watanabe Y, Xu W~R, Yamada T,
  Yamamoto A, Yamamoto K, Yamamoto K, Yamamoto S, Yamamoto T, Yokogawa K,
  Yokoyama J, Yokozawa T, Yoon T~H, Yoshioka T, Yuzurihara H, Zeidler S, Zhu
  Z~H and {KAGRA collaboration} 2019 {\em Nature Astronomy\/} {\bf 3} 35--40
  ISSN 2397-3366
  \urlprefix\url{https://www.nature.com/articles/s41550-018-0658-y}

\bibitem{PuEA2010}
Punturo M, Abernathy M, Acernese F, Allen B, Andersson N, Arun K, Barone F,
  Barr B, Barsuglia M, Beker M, Beveridge N, Birindelli S, Bose S, Bosi L,
  Braccini S, Bradaschia C, Bulik T, Calloni E, Cella G, Mottin E~C, Chelkowski
  S, Chincarini A, Clark J, Coccia E, Colacino C, Colas J, Cumming A,
  Cunningham L, Cuoco E, Danilishin S, Danzmann K, Luca G~D, Salvo R~D, Dent T,
  Rosa R~D, Fiore L~D, Virgilio A~D, Doets M, Fafone V, Falferi P, Flaminio R,
  Franc J, Frasconi F, Freise A, Fulda P, Gair J, Gemme G, Gennai A, Giazotto
  A, Glampedakis K, Granata M, Grote H, Guidi G, Hammond G, Hannam M, Harms J,
  Heinert D, Hendry M, Heng I, Hennes E, Hild S, Hough J, Husa S, Huttner S,
  Jones G, Khalili F, Kokeyama K, Kokkotas K, Krishnan B, Lorenzini M, L{\"u}ck
  H, Majorana E, Mandel I, Mandic V, Martin I, Michel C, Minenkov Y, Morgado N,
  Mosca S, Mours B, M{\"u}ller-Ebhardt H, Murray P, Nawrodt R, Nelson J,
  Oshaughnessy R, Ott C~D, Palomba C, Paoli A, Parguez G, Pasqualetti A,
  Passaquieti R, Passuello D, Pinard L, Poggiani R, Popolizio P, Prato M, Puppo
  P, Rabeling D, Rapagnani P, Read J, Regimbau T, Rehbein H, Reid S, Rezzolla
  L, Ricci F, Richard F, Rocchi A, Rowan S, R{\"u}diger A, Sassolas B,
  Sathyaprakash B, Schnabel R, Schwarz C, Seidel P, Sintes A, Somiya K,
  Speirits F, Strain K, Strigin S, Sutton P, Tarabrin S, Th{\"u}ring A, van~den
  Brand J, van Leewen C, van Veggel M, van~den Broeck C, Vecchio A, Veitch J,
  Vetrano F, Vicere A, Vyatchanin S, Willke B, Woan G, Wolfango P and Yamamoto
  K 2010 {\em Classical and Quantum Gravity\/} {\bf 27} 194002
  \urlprefix\url{http://stacks.iop.org/0264-9381/27/i=19/a=194002}

\bibitem{ET2011}
{ET Science Team} 2011 {\em {available from European Gravitational Observatory,
  document number ET-0106C-10}\/}

\bibitem{ET2020}
{ET Steering Committee} 2020 {\em {available from European Gravitational
  Observatory, document number ET-0007B-20}\/}

\bibitem{ReEA2019}
Reitze D, Adhikari R~X, Ballmer S, Barish B, Barsotti L, Billingsley G, Brown
  D~A, Chen Y, Coyne D, Eisenstein R, Evans M, Fritschel P, Hall E~D, Lazzarini
  A, Lovelace G, Read J, Sathyaprakash B~S, Shoemaker D, Smith J, Torrie C,
  Vitale S, Weiss R, Wipf C and Zucker M 2019 {Cosmic Explorer: The U.S.
  Contribution to Gravitational-Wave Astronomy beyond LIGO} (\textit{Preprint}
  \eprint{1907.04833})

\bibitem{EvEA2021}
Evans M, Adhikari R~X, Afle C, Ballmer S~W, Biscoveanu S, Borhanian S, Brown
  D~A, Chen Y, Eisenstein R, Gruson A, Gupta A, Hall E~D, Huxford R, Kamai B,
  Kashyap R, Kissel J~S, Kuns K, Landry P, Lenon A, Lovelace G, McCuller L, Ng
  K~K~Y, Nitz A~H, Read J, Sathyaprakash B~S, Shoemaker D~H, Slagmolen B~J~J,
  Smith J~R, Srivastava V, Sun L, Vitale S and Weiss R 2021 {A Horizon Study
  for Cosmic Explorer: Science, Observatories, and Community}
  \urlprefix\url{https://arxiv.org/abs/2109.09882}

\bibitem{MaEA2020}
Maggiore M, Broeck C~V~D, Bartolo N, Belgacem E, Bertacca D, Bizouard M~A,
  Branchesi M, Clesse S, Foffa S, Garc{\'{\i}}a-Bellido J, Grimm S, Harms J,
  Hinderer T, Matarrese S, Palomba C, Peloso M, Ricciardone A and Sakellariadou
  M 2020 {\em Journal of Cosmology and Astroparticle Physics\/} {\bf 2020}
  050--050
  \urlprefix\url{https://doi.org/10.1088\%2F1475-7516\%2F2020\%2F03\%2F050}

\bibitem{ChEA2018}
Chan M~L, Messenger C, Heng I~S and Hendry M 2018 {\em Phys. Rev. D\/} {\bf
  97}(12) 123014
  \urlprefix\url{https://link.aps.org/doi/10.1103/PhysRevD.97.123014}

\bibitem{BeEA2011}
Beker M, Cella G, DeSalvo R, Doets M, Grote H, Harms J, Hennes E, Mandic V,
  Rabeling D, van~den Brand J and van Leeuwen C 2011 {\em General Relativity
  and Gravitation\/} {\bf 43} 623--656 ISSN 0001-7701
  \urlprefix\url{http://dx.doi.org/10.1007/s10714-010-1011-7}

\bibitem{Har2019}
Harms J 2019 {\em Living Reviews in Relativity\/} {\bf 22} 6 ISSN 1433-8351
  \urlprefix\url{https://doi.org/10.1007/s41114-019-0022-2}

\bibitem{BaHa2019}
Badaracco F and Harms J 2019 {\em Classical and Quantum Gravity\/} {\bf 36}
  145006 \urlprefix\url{https://doi.org/10.1088%2F1361-6382%2Fab28c1}

\bibitem{AnHa2020}
Andric T and Harms J 2020 {\em Journal of Geophysical Research: Solid Earth\/}
  {\bf 125} e2020JB020401 e2020JB020401 10.1029/2020JB020401
  \urlprefix\url{https://agupubs.onlinelibrary.wiley.com/doi/abs/10.1029/2020JB020401}

\bibitem{BaEA2021}
Bader M, Koley S, van~den Brand J, Campman X, Bulten H~J, Linde F and Vink B
  2021 {\em Classical and Quantum Gravity\/}
  \urlprefix\url{http://iopscience.iop.org/article/10.1088/1361-6382/ac1be4}

\bibitem{BaEA2020}
Badaracco F, Harms J, Bertolini A, Bulik T, Fiori I, Idzkowski B, Kutynia A,
  Nikliborc K, Paoletti F, Paoli A, Rei L and Suchinski M 2020 {\em Classical
  and Quantum Gravity\/} {\bf 37} 195016
  \urlprefix\url{https://doi.org/10.1088/1361-6382/abab64}

\bibitem{AlEA2021}
Allocca A, Berbellini A, Boschi L, Calloni E, Cardello G~L, Cardini A,
  Carpinelli M, Contu A, D'Onofrio L, D'Urso D, Dell'Aquila D, Rosa R~D, Fiore
  L~D, Giovanni M~D, Pace S~D, Errico L, Fiori I, Giunchi C, Grado A, Harms J,
  Majorana E, Mangano V, Marsella M, Migoni C, Naticchioni L, Olivieri M,
  Oggiano G, Paoletti F, Punturo M, Puppo P, Rapagnani P, Ricci F, Rozza D,
  Saccorotti G, Sequino V, Sipala V, Melo I~T~E and Trozzo L 2021 {\em The
  European Physical Journal Plus\/} {\bf 136} 511 ISSN 2190-5444
  \urlprefix\url{https://doi.org/10.1140/epjp/s13360-021-01450-8}

\bibitem{AmEA2020}
Amann F, Bonsignorio F, Bulik T, Bulten H~J, Cuccuru S, Dassargues A, DeSalvo
  R, Fenyvesi E, Fidecaro F, Fiori I, Giunchi C, Grado A, Harms J, Koley S,
  Kovács L, Losurdo G, Mandic V, Meyers P, Naticchioni L, Nguyen F, Oggiano G,
  Olivieri M, Paoletti F, Paoli A, Plastino W, Razzano M, Ruggi P, Saccorotti
  G, Sintes A~M, Somlai L, Ván P and Vasúth M 2020 {\em Review of Scientific
  Instruments\/} {\bf 91} 094504
  \urlprefix\url{https://doi.org/10.1063/5.0018414}

\bibitem{HaHi2014}
Harms J and Hild S 2014 {\em Classical and Quantum Gravity\/} {\bf 31} 185011
  \urlprefix\url{http://stacks.iop.org/0264-9381/31/i=18/a=185011}

\bibitem{SHH2020}
Singha A, Hild S and Harms J 2020 {\em Classical and Quantum Gravity\/} {\bf
  37} 105007 \urlprefix\url{https://doi.org/10.1088/1361-6382/ab81cb}

\bibitem{HaEA2021}
Hall E~D, Kuns K, Smith J~R, Bai Y, Wipf C, Biscans S, Adhikari R~X, Arai K,
  Ballmer S, Barsotti L, Chen Y, Evans M, Fritschel P, Harms J, Kamai B,
  Rollins J~G, Shoemaker D, Slagmolen B~J~J, Weiss R and Yamamoto H 2021 {\em
  Phys. Rev. D\/} {\bf 103}(12) 122004
  \urlprefix\url{https://link.aps.org/doi/10.1103/PhysRevD.103.122004}

\bibitem{SiEA2021}
Singha A, Hild S, Harms J, Tringali M~C, Fiori I, Paoletti F, Bulik T,
  Idzkowski B, Bertolini A, Calloni E, Rosa R~D, Errico L and Gennai A 2021
  {\em Classical and Quantum Gravity\/} {\bf 38} 245007
  \urlprefix\url{https://doi.org/10.1088/1361-6382/ac348a}

\bibitem{DHA2012}
Driggers J~C, Harms J and Adhikari R~X 2012 {\em Phys. Rev. D\/} {\bf 86}(10)
  102001 \urlprefix\url{http://link.aps.org/doi/10.1103/PhysRevD.86.102001}

\bibitem{CoEA2016a}
Coughlin M, Mukund N, Harms J, Driggers J, Adhikari R and Mitra S 2016 {\em
  Classical and Quantum Gravity\/} {\bf 33} 244001
  \urlprefix\url{http://stacks.iop.org/0264-9381/33/i=24/a=244001}

\bibitem{HaEA2020}
Harms J, Bonilla E~L, Coughlin M~W, Driggers J, Dwyer S~E, McManus D~J, Ross
  M~P, Slagmolen B~J~J and Venkateswara K 2020 {\em Phys. Rev. D\/} {\bf
  101}(10) 102002
  \urlprefix\url{https://link.aps.org/doi/10.1103/PhysRevD.101.102002}

\bibitem{PaHa2016}
Paik H~J and Harms J 2016 {\em Journal of Physics: Conference Series\/} {\bf
  716} 012025 \urlprefix\url{http://stacks.iop.org/1742-6596/716/i=1/a=012025}

\bibitem{HaVe2016}
Harms J and Venkateswara K 2016 {\em Classical and Quantum Gravity\/} {\bf 33}
  234001 \urlprefix\url{https://doi.org/10.1088%2F0264-9381%2F33%2F23%2F234001}

\bibitem{HaEA2009b}
Harms J, DeSalvo R, Dorsher S and Mandic V 2009 {Gravity-Gradient Subtraction
  in 3rd Generation Underground Gravitational-Wave Detectors in Homogeneous
  Media} \urlprefix\url{https://arxiv.org/abs/0910.2774}

\bibitem{BeEA2012}
Beker M~G, van~den Brand J~F~J, Hennes E and Rabeling D~S 2012 {\em Journal of
  Physics: Conference Series\/} {\bf 363} 012004
  \urlprefix\url{http://stacks.iop.org/1742-6596/363/i=1/a=012004}

\bibitem{CoEA2018b}
Coughlin M, Harms J, Bowden D~C, Meyers P, Tsai V~C, Mandic V, Pavlis G and
  Prestegard T 2019 {\em Journal of Geophysical Research: Solid Earth\/} {\bf
  124} 2941--2956 (\textit{Preprint}
  \eprint{https://agupubs.onlinelibrary.wiley.com/doi/pdf/10.1029/2018JB016608})
  \urlprefix\url{https://agupubs.onlinelibrary.wiley.com/doi/abs/10.1029/2018JB016608}

\bibitem{CoEA2018a}
Coughlin M~W, Harms J, Driggers J, McManus D~J, Mukund N, Ross M~P, Slagmolen
  B~J~J and Venkateswara K 2018 {\em Phys. Rev. Lett.\/} {\bf 121}(22) 221104
  \urlprefix\url{https://link.aps.org/doi/10.1103/PhysRevLett.121.221104}

\bibitem{TrEA2019}
Tringali M~C, Bulik T, Harms J, Fiori I, Paoletti F, Singh N, Idzkowski B,
  Kutynia A, Nikliborc K, Suchi{\'{n}}ski M, Bertolini A and Koley S 2019 {\em
  Classical and Quantum Gravity\/} {\bf 37} 025005
  \urlprefix\url{https://doi.org/10.1088%2F1361-6382%2Fab5c43}

\bibitem{Naticchioni_2020}
Naticchioni L, Boschi V, Calloni E, Capello M, Cardini A, Carpinelli M, Cuccuru
  S, D'Ambrosio M, de~Rosa R, Giovanni M~D, d'Urso D, Fiori I, Gaviano S,
  Giunchi C, Majorana E, Migoni C, Oggiano G, Olivieri M, Paoletti F, Paratore
  M, Perciballi M, Piccinini D, Punturo M, Puppo P, Rapagnani P, Ricci F,
  Saccorotti G, Sipala V and Tringali M~C 2020 {\em Journal of Physics:
  Conference Series\/} {\bf 1468} 012242
  \urlprefix\url{https://doi.org/10.1088/1742-6596/1468/1/012242}

\bibitem{etsar2}
Di~Giovanni M, Giunchi C, Saccorotti G, Berbellini A, Boschi L, Olivieri M,
  De~Rosa R, Naticchioni L, Oggiano G, Carpinelli M, D’Urso D, Cuccuru S,
  Sipala V, Calloni E, Di~Fiore L, Grado A, Migoni C, Cardini A, Paoletti F,
  Fiori I, Harms J, Majorana E, Rapagnani P, Ricci F and Punturo M 2020 {\em
  Seismological Research Letters\/} {\bf 92} 352--364 ISSN 0895-0695
  (\textit{Preprint}
  \eprint{https://pubs.geoscienceworld.org/ssa/srl/article-pdf/92/1/352/5209839/srl-2020186.1.pdf})
  \urlprefix\url{https://doi.org/10.1785/0220200186}

\bibitem{etterz}
Koley S, Bader M, van~den Brand J, Campman X, Bulten H~J, Linde F and Vink B
  2021 {\em Classical and Quantum Gravity\/}
  \urlprefix\url{http://iopscience.iop.org/article/10.1088/1361-6382/ac2b08}

\bibitem{ET-0426A-21}
Naticchioni L, Saccorotti G, Giunchi C and D'Urso D 2021 {\em available from
  European Gravitational Observatory, document number ET-0426A-21\/}

\bibitem{Pet1993}
Peterson J 1993 {\em Open-file report\/} {\bf 93-322}

\end{thebibliography}

\end{document}